\documentclass[%
reprint,
showpacs,%preprintnumbers,
 amsmath,amssymb,
 aps,
prstab,
]{revtex4-1}

\usepackage{graphicx}% Include figure files
\usepackage{dcolumn}% Align table columns on decimal point
\usepackage{bm}% bold math
\usepackage{color}
\usepackage{todonotes}
\usepackage{hyperref}
\usepackage{siunitx}
\usepackage[caption=false]{subfig}
\usepackage{float}

\usepackage{tikz}
\usetikzlibrary{arrows.meta}
\tikzset{%
    >={Latex[width=2mm,length=2mm]},
    base/.style = {rectangle, rounded corners, draw=black,
        minimum width=3cm, minimum height=1cm,
        text centered},
    input/.style = {base, fill=blue!30},
    output/.style = {base, fill=green!30},
    process/.style = {base, fill=orange!15},
}

\newcommand{\diff}{\mathrm{d}}

\newcommand{\timecoord}{\Delta t}
\newcommand{\energycoord}{\Delta E}
\newcommand{\lineden}{\lambda}
\newcommand{\distrib}{\mathcal{F}}
\newcommand{\hamiltonian}{\mathcal{H}}
\newcommand{\eomfacdE}{A_{\energycoord}}

\newcommand{\potwellrf}{\mathcal{U}_{\mathrm{rf}}}
\newcommand{\potwellimp}{\mathcal{U}_{\mathrm{ind}}}
\newcommand{\action}{\mathcal{J}}
\newcommand{\impedance}{\mathcal{Z}}

\newcommand{\bunchlen}{\tau_l}
\newcommand{\hamillen}{\hamiltonian_l}

\renewcommand{\eqref}[1]{Eq.~(\ref{#1})}
\newcommand{\figref}[1]{Fig.~\ref{#1}}

\begin{document}

%\preprint{APS/123-QED}

\title{Beam Longitudinal Dynamics Simulation Suite BLonD}% Force line breaks with \\

\author{H.~Timko}
\author{S.~Albright}
\author{T.~Argyropoulos} \author{K.~Iliakis}\email{konstantinos.iliakis@cern.ch}
\author{B.~E.~Karlsen-B\ae{}ck}
\author{I.~Karpov}
\author{A.~Lasheen}
\author{L.~Medina}
\author{D.~Quartullo}
\author{J.~Repond}
\affiliation{CERN, Geneva 23, 1211, Switzerland}

\author{J.~Esteban M\"uller}
\affiliation{European Spallation Source, Lund, Sweden}

\author{M.~Schwarz}
\affiliation{Karlsruhe Institute of Technology, Karlsruhe, Germany}

\author{P.~Tsapatsaris}
\author{G.~Typaldos}
\affiliation{National Technical University of Athens, Athens, Greece}

\date{\today}% It is always \today, today,
             %  but any date may be explicitly specified

\begin{abstract}

The beam longitudinal dynamics code BLonD has been developed at CERN since 2014 and has become a central tool for longitudinal beam dynamics simulations. In this paper, we present this modular simulation suite and the various physics models that can be included and combined by the user. We detail the reference frame, the equations of motion, the BLonD-specific options for radio-frequency parameters such as phase noise, fixed-field acceleration, and feedback models for the CERN accelerators, as well as the modeling of collective effects and synchrotron radiation. We also present various methods of generating multi-bunch distributions matched to a given impedance model. BLonD is furthermore a well-tested and optimized simulation suite, which is demonstrated through examples, too.

%\begin{description}
%\item[PACS numbers] %May be entered using the \verb+\pacs{#1}+ command.
%\end{description}
\end{abstract}

\pacs{29.20.D, 29.27.-a, 07.05.Tp}% PACS, the Physics and Astronomy
                             % Classification Scheme.
% 29.20.D - Accelerators, cyclic
% 07.05.Tp - Computer modelling and simulation
% 29.27.-a - Beams, charged particle in accelerators
%\keywords{Suggested keywords}%Use showkeys class option if keyword
                              %display desired

\maketitle

%\tableofcontents

%%% INTRODUCTION %%%
\section{\label{sec:intro} Introduction}

For several decades, longitudinal beam dynamics simulations at CERN have been performed using Fermilab's widely-used simulation suite called ESME~\cite{MacLachlan:1984}. With the upgrade projects of the CERN synchrotrons and studies of future machines, there was a growing need for precision simulations that can combine for a given study all relevant physics with machine-specific features. At the same time, the simulation suite would have to be general enough to cover a wide range of applications, from low- to high energy synchrotrons, from electrons over protons to ions, from space-charge to synchrotron-radiation dominated regimes.

The Beam Longitudinal Dynamics simulation suite BLonD~\cite{BLonD-website,BLonD-github} has been developed at CERN since 2014. It is an open-source particle tracking code for simulation of longitudinal motion in synchrotrons, written in Python and C\texttt{++} languages. It relies on some of the most popular and efficient scientific libraries including Numpy~\cite{walt2011numpy}, and Scipy~\cite{jones2014scipy}. 

The code uses macro-particles with the same charge-to-mass ratio as the real particles. Real bunch intensities typically vary in the range of $10^8-10^{13}$~particles/bunch, while in simulations, an order of $10^4-10^7$~macro-particles/bunch are used. BLonD has a modular structure that allows the user to model different effects according to his/her needs. The BLonD code's unique feature is that it disentangles the equations of the beam particles and the radio-frequency (rf) system, and tracks both of them with respect to an external reference `clock', just as in a real machine. This has the advantage of being able to include several beam- and/or cavity loops, collective effects, etc.\ when modeling the beam motion. Additional special features of the code are the generation of matched (multi-)bunch phase-space distributions with collective effects, rf phase noise and sinusoidal rf phase modulation, and its overall flexibility.

%Present applications.
By now, the BLonD suite has been thoroughly benchmarked~\cite{Timko:2016} and applied for all the CERN existing and future synchrotrons; the outcome of the simulation studies has often been guiding the baseline choices for machine upgrades and studies. Several other institutes have started to use the code as well~\cite{Ohmori2017,Geithner2019}. To mention a few applications, the CERN Proton Synchrotron Booster (PSB), for instance, poses its challenges with an injection scheme using constant frequency during a magnetic ramp, with beam-based feedback systems on, strong space-charge, and a beam that stretches over the entire ring before getting bunched. The Proton Synchrotron (PS) is equipped with numerous rf systems used to shape the beam via rf manipulations (e.g.\ splitting and merging bunches), so simulations including multi-rf systems and beam loading effects are required to predict the best achievable beam quality when increasing beam intensity. Determining the beam and rf parameters for the post-2021 Super Proton Synchrotron (SPS) requires a meticulously accurate impedance model, which was determined from benchmarks of beam-based measurements against simulation results. The successful operation of the Large Hadron Collider (LHC) relies on controlled emittance blow-up performed during a 14-million-turn energy ramp with several feedback loops interacting. The proposed electron-positron Future Circular Collider (FCC) poses challenges with its numerical requirements for resolving small beams in a large machine. 

In this article, we present the unique features and special applications of the BLonD code. The paper is structured as follows: starting from the (i) reference frame and beam-cavity interactions, we then describe (ii) the modulation of rf parameters, (iii) the modeling of impedance and collective effects, (iv) synchrotron radiation and quantum excitation, (v) global and local feedback models, over the (vi) generation of particle distributions, to (vii) optimizations and (vii) benchmark techniques.

%%% REFERENCE FRAME %%%
\section{\label{sec:refframe} Reference frame and beam-cavity interactions}

Just like beam- and cavity control work in a real synchrotron, BLonD describes the evolution of beam particle coordinates and the voltage, phase, and frequency in the radio-frequency accelerating cavities w.r.t.\ an external reference clock. The design clock time $t_{d,(n)}$ in a given turn $n$ counts the total number of turns elapsed in the laboratory frame,
\begin{equation}
  \label{eq:referenceTime}
  t_{d,(0)} \equiv 0 \phantom{xx}\text{and}\phantom{xx} t_{d, (n)} \equiv \sum_{k=1}^n T_{\mathrm{rev},(k)} \phantom{x}\mathrm{for}\phantom{x} n \geq 1 .
\end{equation}
The revolution periods $T_{\mathrm{rev},(n)}$ are defined by the design orbit of radius $R_d$ and $\beta_{d,(n)}$, the relative speed of the design particle with respect to the speed of light $c$ on that orbit,
\begin{equation}
  \label{eq:revolutionPeriod}
  T_{\mathrm{rev},(n)} = \frac{2 \pi R_d}{\beta_{d,(n)}c} ,
\end{equation}
where the user can input $\beta_{d,(n)}$ implicitly via the corresponding design relativistic momentum $p_{d,(n)}$ or total energy $E_{d,(n)}$ evolution over time. The input also determines the design magnetic field ramp $B_{d,(n)}$ through the relation
\begin{equation}
  \label{eq:magneticField}
  p_{d,(n)} = |q| \rho B_{d,(n)},
\end{equation}
where $q$ is the charge of the particle and $\rho$ the bending radius of the magnets.

\begin{figure}[!ht]
  \includegraphics[width=0.35\textwidth]{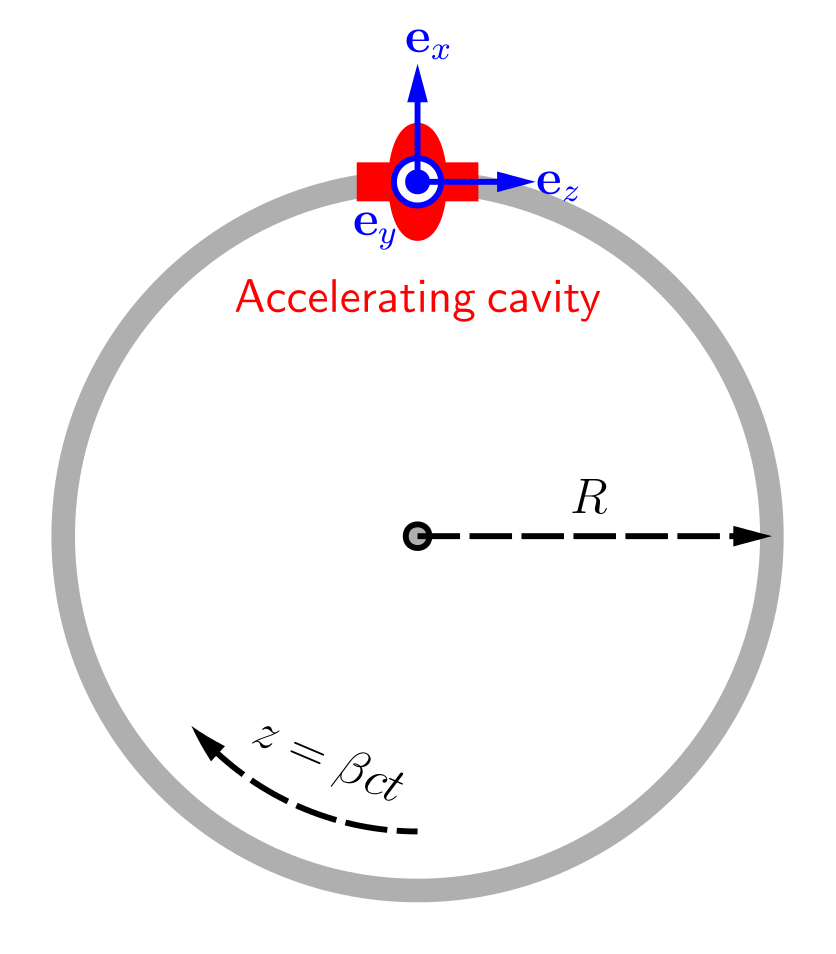}
  \caption{Reference frame for the equations of motion and the reference time. A given particle on the orbit $R$ is described with the energy $E$ and the longitudinal coordinate $t$.}
  \label{fig:coordinateSystem}
\end{figure}

The user can choose to place several rf stations along the ring, in which case the magnetic field program of each section of the ring (from one rf station to another) has to be input. This sub-cycling of a turn can be useful, for instance, in the presence of strong synchrotron radiation. In each rf station, an arbitrary number of $n_\mathrm{rf}$ harmonic rf systems can be modeled; all rf systems in a given rf station will be treated as having the same longitudinal position. The origin of the coordinate system $(e_x, e_y, e_z)$ is fixed to the longitudinal position of the first rf station, on the reference orbit, see Fig.~\ref{fig:coordinateSystem}.

The particles are described with the phase space coordinates $(\Delta t_{(n)}, \Delta E_{(n)})$, which are the particle's arrival time and energy with respect to the integrated reference time $t_{d,(n)}$ and the design total energy $E_{d,(n)}$, respectively. For each section of the ring, first the energy $\Delta E$ of a given particle is updated from time step $n$ to $n+1$, based on the particle's arrival time and all the rf voltage kicks $k$ received in the corresponding rf station,
\begin{align}
  \Delta E_{(n+1)} &=  \Delta E_{(n)} + \sum_{k=1}^{n_{\mathrm{rf}}} q V_{k,(n)} \sin(\omega_{\mathrm{rf},k,(n)} \Delta t_{(n)} + \varphi_{\mathrm{rf},k,(n)}) - \nonumber \\
  &- (E_{d,(n+1)} - E_{d,(n)}) + E_{\mathrm{other},(n)} ,
  \label{eq:kick}
\end{align}
where $V_k$ is the voltage amplitude, $\omega_{\mathrm{rf},k}$ the rf angular frequency, and $\varphi_{\mathrm{rf},k}$ the phase of the rf system $k$, and $E_{d,(n+1)} - E_{d,(n)}$ the change of the design energy from one turn to another. The last term $E_{\mathrm{other},(n)}$ contains additional energy changes due to induced voltage, synchrotron radiation, etc. The time coordinate of the particle is updated subsequently, using the already updated energy of the particle and the momentum compaction factor $\alpha$ of at least zeroth, and up to second order,
\begin{align}
  \Delta t_{(n+1)} &= \Delta t_{(n)} + T_{\mathrm{rev},(n+1)}\Bigg[  \Big( 1 + \alpha_{0,(n+1)} \delta_{(n+1)}  + \nonumber \\
  &+ \alpha_{1,(n+1)} \delta^2_{(n+1)} + \alpha_{2,(n+1)} \delta^3_{(n+1)} \Big) \frac{1 + \frac{\Delta E_{(n+1)}}{E_{d,(n+1)}}}{1 + \delta_{(n+1)}} - 1 \Bigg] ,
  \label{eq:drift}
\end{align}
where $\delta_{(n)} = \frac{\Delta p_{(n)}}{p_{d,(n)}}\simeq\frac{\Delta E_{(n)}}{\beta_d^2 E_{d,(n)}}$ is the relative momentum offset.

\subsection{Periodicity}

\begin{figure}[!htb]
  \includegraphics[width=0.4\textwidth]{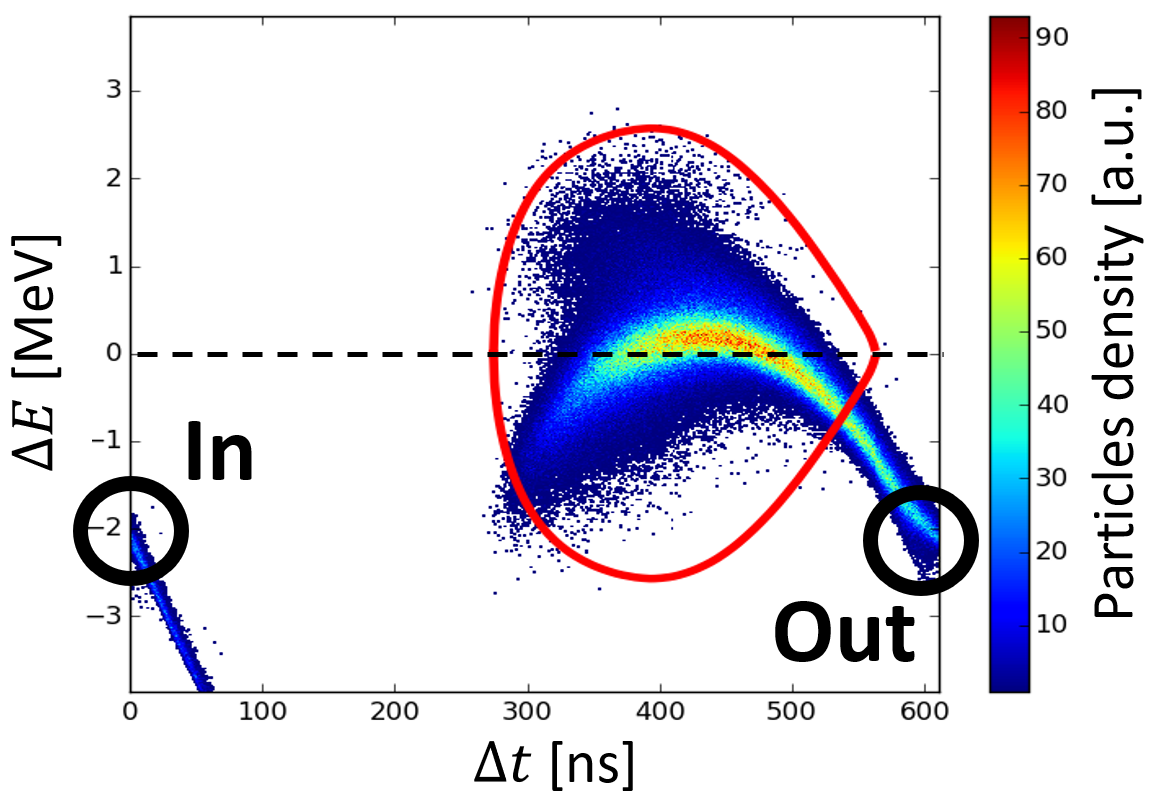}
  \caption{Phase-space plot of unstable bunch in accelerating bucket ($h=1$). The time coordinate on the horizontal axis spans from 0 to $T_{\mathrm{rev}}$. With the periodic condition enabled, particles that exit the time frame from the right enters back into the frame from the left.}
  \label{fig:periodicity}
\end{figure}

In some specific cases, such as the PSB operated at the $h=1$ harmonic, a single bunch can cover the entire machine circumference, with the head and the tail of the bunch crossing the boundaries between previous, present, and next turns, see Fig.~\ref{fig:periodicity}. 
For such cases, the periodicity of the equations of motion can be enabled. The periodicity will ensure that all particles remain synchronized to the present time frame, by tracking twice the particles that lag behind, and by pausing the tracking of particles that are ahead in time.

If the periodicity algorithm is not applied, the equations of motion make all the particles that cross the line $\Delta t = T_{\mathrm{rev}}$ in phase space drift away from the rf bucket. This ignores the geometry of the ring assumed in the equations of motion, and could give undesirable results for example when the beam losses have to be computed accurately or when an un-bunched beam has to be captured inside an rf bucket.

The periodicity algorithm is also useful when parts of the rf bucket cross the line $\Delta t = T_{\mathrm{rev}}$ in phase space, for example when the beam phase loop changes the design rf frequency and shifts the bucket in phase space, or when a second rf system
with relatively high voltage is added in bunch-lengthening mode to the main-harmonic voltage during acceleration. In these cases, the bunch will be numerically split in phase space into two portions, which however in practice behave as a whole.

\section{Modulation of RF parameters}

BLonD provides the possibility of simulating any complex beam manipulation in the longitudinal plane, since both the momentum and the rf programs ($\omega_{\mathrm{rf},k}(t)$, $V_{k}(t)$, $\varphi_{\mathrm{rf},k}(t)$) can be given as an input. Below, we show some examples on rf phase noise and modulation, as well as slip stacking. However, many other applications, such as cogging, synchronization, fixed-frequency acceleration, etc.\ are possible to simulate, too.

\subsection{RF phase noise and modulation}

For controlled longitudinal emittance blow-up~\cite{Krinsky:1982,Dome:1985}, both band-limited rf phase noise of the main harmonic~\cite{Timko:2015,Albright_2019} and single-frequency modulation of a high harmonic~\cite{Shaposhnikova:2014} can be used. In BLonD, a turn-by-turn phase offset ($\Delta \varphi_{\mathrm{rf},(n)}$) can be added to the programmed rf phase ($\varphi_{\mathrm{rf},(n)}$) to achieve this. The modulation functions can be defined by the user directly, or calculated with built-in functions. For instance, the generation method of the phase noise presently used in the CERN synchrotrons~\cite{Tuckmantel:2008} can be applied.

In the case of single-frequency modulation, $\Delta\varphi$ is computed as
\begin{equation}
    \Delta\varphi_{n} = A\sin\left(2\pi\int{f_n \diff t}\right) + \varphi_{\mathrm{off}}
\end{equation}
where $A$ is the modulation amplitude, $f_n$ is the modulation frequency and $\varphi_{\mathrm{off}}$ is an offset about which the modulation is applied.  To correctly simulate the phase modulation a frequency modulation is computed at the same time, defined by:
\begin{equation}
    \Delta\omega_n = \frac{1}{2\pi}\frac{d\Delta\varphi_n}{dt}\omega_n \, .
\end{equation}

\subsection{Fixed-field manipulations: slip stacking}

\begin{figure*}[!htb]
\centering
\includegraphics*[width=0.75\textwidth]{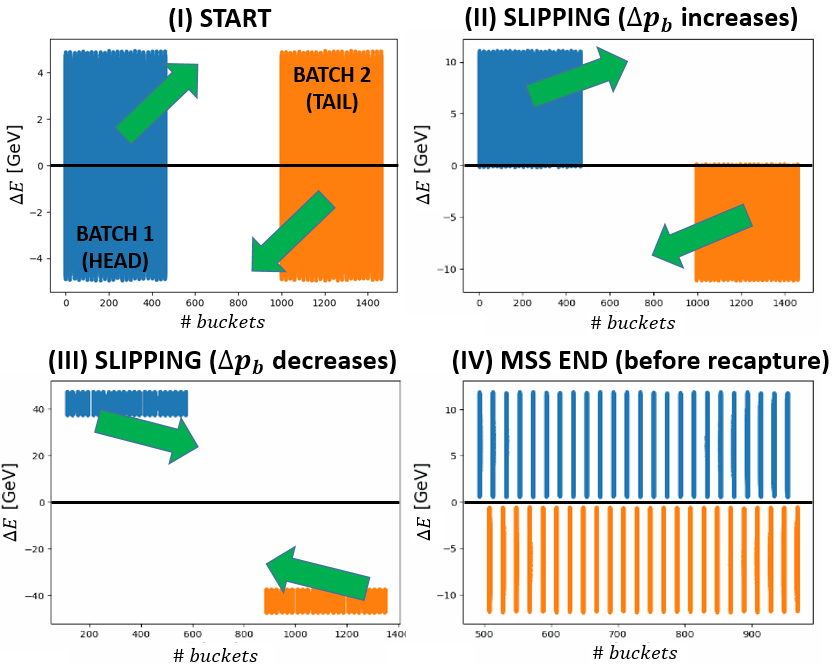}
\caption{Illustration of the MSS procedure simulated in BLonD. Two batches, starting from Phase~I (top left) are separated in energy and move in the longitudinal phase space relative to each other. The black line marks $\Delta E = E-E_{\mathrm{d}} = 0$. In Phase~II (top right), the energy distance between the batches increases, while the opposite occurs in Phase~III (bottom left). Recapture is done in Phase~IV (bottom right).}
\label{fig:MSS_scheme}
\end{figure*}

Momentum slip-stacking (MSS) is one of the most complicated rf manipulations and is currently being studied in simulations for the ion beams in the SPS~\cite{LIU-CERN} using the BLonD code. It permits two high-energy particle beams of slightly different momenta to slip azimuthally, relative to each other, in the same beam pipe. The two beams are captured by two rf systems with a small frequency difference between them. Each beam is synchronized with one rf system and it is perturbed by the other. When the two beams are in the desired azimuthal position, the full beam is recaptured with a much higher rf voltage at the design rf frequency. In particular, for the SPS, two batches of 24 bunches, spaced by 100~ns, are interleaved on an intermediate momentum plateau to produce a single batch of 48 bunches with half the bunch distance (50~ns). The process as simulated in BLonD, is schematically illustrated in Fig.~\ref{fig:MSS_scheme}. Details on how MSS is going to be applied in the SPS can be found in~\cite{DaniloThesis,IPAC19MSS}. 

During MSS the magnetic field $B_d$ will be constant, which means that,  in a first-order approximation, the following relation holds 
\begin{equation}\label{eq:differential_quantities}
\frac{\Delta \omega_{\mathrm{rf}}}{\omega_{\mathrm{rf,d}}} = - \eta_0 \frac{\Delta p}{p_{\mathrm{d}}},
\end{equation}
where $\Delta \omega_{\mathrm{rf}}$ and $\Delta p$ are the changes of the rf angular frequency and beam momentum with respect to their designed values. In a reference frame that is synchronized with the design revolution period $T_{\mathrm{rev,d}}$, a variation $\Delta \omega_\mathrm{rf}$ implies a change in the rf phase according to
\begin{equation}\label{eq:phase_change}
\Delta \varphi_{\mathrm{rf}} = \frac{2 \pi h \Delta \omega_{\mathrm{rf}}}{\omega_{\mathrm{rf,d}}}.
\end{equation}

Providing as an input to BLonD the aforementioned programs, as well as the rf voltage amplitude programs for the two cavity groups, the total rf voltage experienced by each beam is given by
\begin{equation}\label{eq:total_voltage}
V_{\mathrm{rf}} = \hat{V}_{\mathrm{rf},1}\sin(\omega_{\mathrm{rf},1}t+\phi_{\mathrm{rf},1})+\hat{V}_{\mathrm{rf},2}\sin(\omega_{\mathrm{rf},2}t-\phi_{\mathrm{rf},2}),
\end{equation}
where the indices 1,2 indicate the first and second group of rf cavities accordingly.

%%% INTENSITY EFFECTS %%%
\section{\label{sec:impedance} Modeling impedance and collective effects}
In particle accelerators with high-intensity and high-brightness beams, the performance is usually limited by the electromagnetic interaction of the particles with each other and their surroundings, which are known as collective effects.

An outcome of these effects is known as potential-well distortion, which causes, on one hand, incoherent effects as a shift in the synchronous phase and in the synchrotron frequency, as well as bunch lengthening~\cite{Laclare:611596}. For example, the synchrotron frequency shift may affect the performance of a controlled longitudinal emittance blow-up~\cite{Quartullo_2017}. On the other hand, coherent effects can trigger instabilities, which can affect individual bunches or couple several bunches, and can produce uncontrolled emittance blow-up and high beam losses~\cite{Sacherer73}.

BLonD can accurately simulate these effects using the concept of wakefields and impedances~\cite{chao1993physics}. The electromagnetic interaction is described by the so-called wake function $W(t)$, which represents the electric field excited by a point charge as experienced by a test charge.

Computing the interaction between the individual macroparticles is a $O(N^2)$ problem. In most practical cases, it is desirable to apply a binning to the particle distribution and calculate the wake potential, also called induced voltage $V_\mathrm{ind}(\Delta t)$, which is defined as the convolution of the line density $\lambda (t)$ of the beam, normalized to $\int_{-\infty}^{+\infty} \lambda(t) \, \diff t = 1$, and the wake function:
\begin{equation}
V_\mathrm{ind}(\Delta t) = - q\, N_p\int_{-\infty}^{+\infty} \lambda(\tau) \, W(\Delta t-\tau) \, \diff \tau  \, ,\label{eq:Vind_time}
\end{equation}
where $N_p$ is the number of real particles in the beam.

The energy kick $E_\mathrm{ind}(\Delta t) = q\, V_\mathrm{ind}(\Delta t)$ due to the induced voltage enters the term $E_{\mathrm{other}}$ in~\eqref{eq:kick}, where the minus sign in~\eqref{eq:Vind_time} ensures an energy loss for a positive wake potential.

The induced voltage can also be computed in frequency domain using the concept of impedance $Z(\omega)$, defined as the Fourier transform of the wake function:

\begin{equation}
Z(\omega) = \int_{-\infty}^{+\infty} W(t) \, e^{-j \, \omega \, t} \, \diff t ,
\end{equation}
\begin{equation}
V_\mathrm{ind}(\Delta t) = - \frac{ q\, N_p}{2\,\pi} \int_{-\infty}^{+\infty} Z(\omega) \,  \Lambda(\omega)\,e^{j \, \omega \, \Delta t} \, \diff\Delta t ,\label{eq:Vind_freq}
\end{equation}
where $\Lambda(\omega)$ is the beam spectrum, obtained as the Fourier transform of the line density.

Although both time- and frequency-domain methods are equivalent, the discretization required for numerical simulations makes each method suitable for different situations, depending on the bandwidth of the impedance source.

Wakefields corresponding to narrow-band impedance sources require very high frequency resolution to be able to resolve it correctly. However, in time domain we only need to describe the signal for a time window equivalent to the length of a bunch or a train of bunches. In that case, computations in time domain will require significantly less resources. Alternatively, a broadband impedance will result in a wakefield that would require too fine resolution in time domain. 

For these reasons, BLonD implements both time and frequency-domain calculations, as well as different strategies to deal with the discretization of the line density.

\subsection{Impedance sources}

Wake functions and impedances can be calculated analytically for simple geometries (see e.g.,~\cite{Zotter1998}) or using numerical codes for more complicated devices.

BLonD includes several analytical impedance models. One of them is the resonator model, with an impedance defined as
\begin{equation}
Z(f) = \frac{R_s}{1 + j Q \left(\frac{f}{f_r}-\frac{f_r}{f}\right)},
\end{equation}
where $R_s$ is the shunt impedance, $f_r$ is the resonant frequency, and $Q$ the quality factor. The user can include one or more resonators to model the impedance of an element.

Furthermore, BLonD contains a model for the resistive wall impedance of the beam pipe, for the case of a cylindrical beam pipe, and a model for traveling wave cavities. Coherent synchrotron radiation can be modeled as an impedance, too. 

Finally, BLonD also takes as input tables that consist of either time and wakefield values, or frequency and impedance values, which can be used to describe an arbitrary impedance model.

All of the above-mentioned impedance models can be used in a combined manner as well. For more details, see BLonD documentation.

\subsection{Induced voltage calculation}
BLonD uses the impedance sources described above to calculate the induced voltage that is then applied to the macroparticles in the same way as the voltage kick from the accelerating cavities. Both operations can be combined for optimization by summing the rf voltage to the induced voltage with the same time resolution. The total energy kick is then linearly interpolated and applied based on the particle position $\Delta t_i$.

BLonD implements two generic algorithms to compute the induced voltage that works with all the impedance sources described above; one is in time domain and the other in frequency domain.

In frequency domain, the method consists of discretizing Eq.~(\ref{eq:Vind_freq}) as
\begin{equation}
V_\mathrm{ind}[n]  = - q \, N_p \, \mathrm{IDFT}\left( Z[k] \ \Lambda [k]\right),\label{eq:Vind_discr_f}
\end{equation}
where $\mathrm{IDFT}$ is the Inverse Discrete Fourier Transform and $\Lambda [k]$ is the Discrete Fourier Transform ($\mathrm{DFT}$) of the line density: $ \Lambda [k] = \mathrm{DFT}\left( \lambda [n]\right)$.

In time domain, the induced voltage is calculated as a discrete convolution. However, it is in general more efficient to compute the discrete convolution using the DFT according to the circular convolution theorem. \begin{equation}
V_\mathrm{ind}[n]  = - q \, N_p \, \mathrm{IDFT}\left\{ \mathrm{DFT}\left( W[n] \right) \, \mathrm{DFT}\left( \lambda [n]\right)\right\}.\label{eq:Vind_discr_t}
\end{equation}
It is important to carefully pad the two signals with zeros so that the result is the linear convolution. If the length of $W[n]$ is $N$ and the length of $\lambda [n]$ is M, both signals need to be padded so that their length is at least $L = N+M -1$; in practice, the next regular number is used for runtime efficiency. In this case, the complexity of the algorithm using FFT is $O(L\log{}L)$, to be compared to $O(N M)$ using the direct convolution algorithm in time domain.

Given the similarities between Eq.~(\ref{eq:Vind_discr_f}) and Eq.~(\ref{eq:Vind_discr_t}), BLonD has a single implementation for both methods, with the difference that when doing time-domain calculations, a pseudo-impedance is defined as $Z^*[k] = \mathrm{DFT}\left( W^*[n] \right)$, where the * represents that the signal is zero-padded. Similarly, the beam spectrum is defined as $\Lambda^*[k] = \mathrm{DFT}\left( \lambda^*[n] \right)$.

For the special case of purely imaginary impedances, another method is available, in which the magnitude of the impedance is directly proportional to the frequency. In this case, the impedance is described by a constant number, $Z/n$, where $n=\omega/\omega_\mathrm{rev}$. The induced voltage computation is much simpler, as the DFTs are replaced by a derivative:
\begin{equation}
V_\mathrm{ind}[n] = -\frac{q\, T_\mathrm{rev}}{2 \, \pi\, T_s} \frac{Z}{n}  \frac{d\lambda [n]}{dn},
\end{equation}
where $T_s$ is the binning size for the discretized line density.  An interesting use of this method is the modeling of space charge as a reactive impedance (see e.g.,~\cite{Wang:2014}).

BLonD implements also two special algorithms that are limited to resonator impedances. The MuSiC algorithm~\cite{PhysRevSTAB.18.031001} uses a propagation matrix to compute the induced voltage without binning. The second algorithm is an adaptation of a semi-analytic method~\cite{PhysRevSTAB.17.050701} to resonators and does not require uniform binning. 

BLonD can also take into account wakefields lasting more than one turn, which can be an important contribution of narrow-band impedance sources in small rings. The code keeps in memory the induced voltage generated for a user-defined number of preceding turns, which are shifted in time after every turn and added to the induced voltage of the current turn. When the revolution frequency is not constant, a linear interpolation is done to compute the induced voltage with the right binning.

All these algorithms can be combined to represent a full machine impedance model, using each of them for different impedance sources depending on their characteristics. BLonD finally calculates the induced voltage as the sum of the induced voltage from each impedance source, which is then applied to the macroparticles.

%%% SYNCHROTRON RADIATION %%%
\section{\label{sec:synchrad} Synchrotron radiation and quantum excitation}

BLonD is not only used for studying the beam dynamics of existing accelerators, but also for designing future machines. The Future Circular Collider (FCC) project considers two main options that can be installed in a tunnel of about 100~km: a lepton machine (FCC-ee) operating with up to 365~GeV collision energy~\cite{FCCee2019}, and a 100~TeV hadron machine (FCC-hh)~\cite{FCChh2019}. For both of them, synchrotron radiation becomes an essential part of the beam dynamics and needs to be included in macro-particle simulations. This is also true for light sources, which use electron beams to provide X-rays to user experiments.

The synchrotron radiation module in BLonD applies the energy kicks due to the average energy loss per particle per turn $U_0$, a damping term proportional to the particle energy offset, and quantum excitation. The detailed implementation is summarized in Ref.~\cite{JEMuller2017}. Later, the option of an `empty rf station' was added, which applies all kicks except for the rf kick. This is especially important for FCC-ee t$\mathrm{\bar{t}}$, which has a collision energy of 365~GeV and $U_0 = 9.2$~GeV. Its double rf system is distributed in two opposite locations of the ring and consists of multi-cell cavities operating at 400~MHz and 800~MHz providing 4~GV and 6.9~GV of rf voltage, respectively.

\begin{figure}[!ht]
  \centering
  \includegraphics[width=0.47\textwidth]{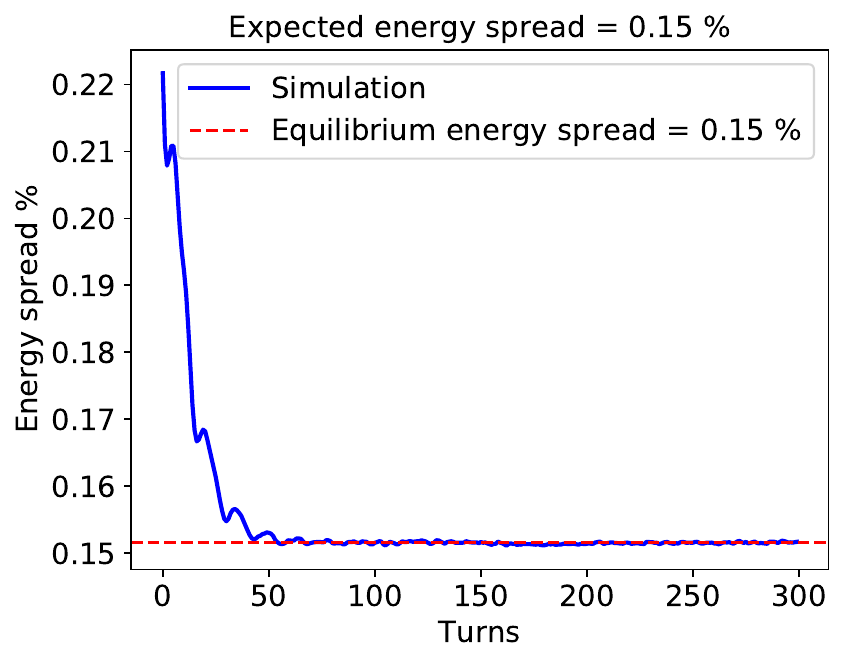}
  \caption{Evolution of the normalized rms energy spread in FCC-ee t$\mathrm{\bar{t}}$ with 365~GeV collision energy. Simulation and machine parameters: the two rf stations contain a 400~MHz rf system with 4~GV rf voltage and 800~MHz rf system with 6.9~GV rf voltage in total; six rf stations are empty.}
  \label{fig:FCC-ee}
\end{figure}

The importance of a correct distribution of kicks can be understood through a simulation example, in which a single bunch is generated with a normalized relative rms energy spread of about $2.2\times 10^{-3}$, using eight rf stations, out of which only two provide half of the total voltage each. The others are `empty' and activated by setting zero rf voltage for them. The evolution of the normalized rms energy spread shows that the equilibrium value agrees well with the analytic expectation~\cite{SYLee}, see Fig.~\ref{fig:FCC-ee}. In similar simulations with only two rf stations, and without additional empty rf stations, particles are lost from the bucket due to large discrete kicks given by synchrotron radiation and quantum excitation. 

%%% FEEDBACKS %%%
\section{Global and local feedback models}

In this Section, we will describe global and local feedback models. By global feedback models, we mean models that are acting on the entire rf system, i.e.\ either the rf frequency or phase. By local feedback models, we mean models that act on one single rf system, typically the rf voltage amplitude and phase of a group of cavities. Often, the feedback models also involve measuring beam signals and feeding the signals back or forward to regulate the desired quantity.

\subsection{Global feedback models}

For accurate and realistic modeling of beam motion in a synchrotron, the presence of beam-based feedback systems cannot always be neglected. For example, quantitatively reproducing measured capture losses in the SPS requires the inclusion of the SPS beam phase loop in the simulation. Also, rf phase noise or modulation is in practice often injected through the set point of a beam phase loop, instead of being injected directly into the cavity set point.

The BLonD equations of motion allow the actual turn-by-turn rf phase $\varphi_{\mathrm{rf},(n)}$ and frequency $\omega_{\mathrm{rf},(n)}$ to deviate from the originally designed phase $\varphi_{\mathrm{rf,d},(n)}$ and frequency $\omega_{\mathrm{rf,d},(n)}$ programs. This feature enables the user to dynamically change the rf parameters throughout the simulation, or simply to program an rf frequency that is not an integer multiple of the revolution frequency of the clock. The deviations of the rf frequency from the exact multiple of the design revolution frequency will result in an accumulated phase deviation,
\begin{equation}
  \label{eq:phaseShift}
  \Delta\varphi_{\mathrm{rf},(n)} = \sum_{k=1}^n \frac{\omega_{\mathrm{rf},(k)}}{h_{(k)} \omega_{\mathrm{rev},(k)}} 2 \pi h_{(k)} .
\end{equation}
In such a case, the user will see the bunch and the rf bucket drift with respect to the reference frame. Frequency loops can be applied to minimize this phase drift and to ensure that in the long run the rf frequency (and with it, the beam orbit and energy) is maintained at its design value.

The exact implementations of the beam-based feedback models are machine-specific. For the CERN synchrotrons, the exact implementations of frequency, synchronization, radial, and beam phase loop are available for use. They can be a good starting point also for feedback loops in other machines, while also custom-made feedback models can be easily built and used with the BLonD architecture.

\subsection{Cavity feedback models}

In the presence of beam and due to beam loading~\cite{Boussard:167557}, both the amplitude and phase of the rf voltage deviate from their design values that are usually constant over a tracking turn. Given that the rf cavities are often the largest contributor to the machine impedance, the correction needed to bring these parameters back to their design values can be computed as in the real machine, by a cavity-based feedback system. Depending on the system, the correction is applied within the same turn for fast feedback systems or with a one-turn delay.

\begin{figure*}[t!b]
    \centering
    % \includegraphics*[width=0.95\textwidth]{SPS_cavitycontrol}
    %\small
  \begin{tikzpicture}[thick,scale=1, every node/.style={scale=0.70}]
    \pgfmathsetmacro{\size}{5.0} % 5.0 is equal to the page width, with the box W/H sizes below
    \pgfmathsetmacro{\xposzero}{(0.000*\size}
    \pgfmathsetmacro{\yposzero}{(0.000*\size}
    \pgfmathsetmacro{\xstep}{0.140*\size} % Boxes width W, relative to size
    \pgfmathsetmacro{\ystep}{0.200*\size} % Boxes height H, relative to size
    \pgfmathsetmacro{\yrowspace}{2.75*\ystep}
    % Open loop
    \pgfmathsetmacro{\xpos}{(1.000*\xposzero}
    \pgfmathsetmacro{\ypos}{(1.000*\yposzero}
    \node at (\xpos+1.50*\xstep, \ypos-0.25*\ystep) [black]  {@$f_\text{rf}$};
    \node at (\xpos+1.50*\xstep, \ypos-0.55*\ystep) [black,font=\scriptsize]  {(carrier)};
    \draw[black, thick, -] (\xpos, \ypos) -- (\xpos+0.75*\xstep, \ypos) -- (\xpos+1.25*\xstep, \ypos+0.25*\ystep) node[draw,circle,fill,scale=0.05*\size] {};
    \node at (\xpos+1.00*\xstep, \ypos+0.50*\ystep) [black,font=\scriptsize]  {open loop};
    \draw[black, thick, ->] (\xpos+1.25*\xstep, \ypos) node[draw,circle,fill,scale=0.05*\size] {} -- (\xpos+2.0*\xstep, \ypos);
    % Sum ant+sp
    \pgfmathsetmacro{\xpos}{\xpos+2.0*\xstep}
    \pgfmathsetmacro{\xsumantsp}{\xpos+0.5*\xstep}
    \node at (\xsumantsp, \ypos) [black] {$\sum$};
    \draw[thick] (\xsumantsp, \ypos) circle (0.5*\xstep);
    \node at (\xpos+0.0*\xstep, \ypos+0.0*\xstep) [black,anchor=south east] {$-$};
    \node at (\xpos+0.5*\xstep, \ypos+0.5*\ystep) [black,anchor=west] {$+$};
    \node at (\xpos+1.5*\xstep, \ypos) [black,anchor=south] {$\Delta \vec{V}$};
    \draw[black, thick, ->] (\xpos+1.0*\xstep, \ypos) -- (\xpos+2.0*\xstep, \ypos);
    % Gfb
    \pgfmathsetmacro{\xpos}{\xpos+2.0*\xstep}
    \node at (\xpos+0.375*\xstep, \ypos) [black] {$G_\text{fb}$};
    \draw[thick] (\xpos, \ypos-0.50*\ystep) -- (\xpos, \ypos+0.50*\ystep) -- (\xpos+\xstep,\ypos) -- (\xpos, \ypos-0.50*\ystep);
    \draw[black, thick, ->] (\xpos+1.0*\xstep, \ypos) -- (\xpos+1.5*\xstep, \ypos);
    % Comb
    \pgfmathsetmacro{\xpos}{\xpos+1.5*\xstep}
    \node at (\xpos+0.50*\xstep, \ypos) [black]  {$H_\text{comb}$};
    \draw[thick] (\xpos, \ypos-0.50*\ystep) rectangle (\xpos+1.0*\xstep, \ypos+0.50*\ystep);
    \draw[black, thick, ->] (\xpos+1.0*\xstep, \ypos) -- (\xpos+2.0*\xstep, \ypos);
    % Mod
    \pgfmathsetmacro{\xpos}{\xpos+2.0*\xstep}
    \node[rotate=90]  at (\xpos+0.25*\xstep, \ypos) [black] {mod};
    \draw[thick] (\xpos, \ypos-0.50*\ystep) rectangle (\xpos+0.50*\xstep, \ypos+0.50*\ystep);
    \node at (\xpos-0.50*\xstep, \ypos-0.25*\ystep) [black]  {@$f_\text{rf}$};
    \node at (\xpos+1.00*\xstep, \ypos-0.25*\ystep) [black]  {@$f_\text{r}$};
    \node at (\xpos+1.00*\xstep, \ypos-0.70*\ystep) [black,font=\scriptsize]  {(central resonant)};
    \draw[black, thick, ->] (\xpos+0.5*\xstep, \ypos) -- (\xpos+1.5*\xstep, \ypos);
    % Delay
    \pgfmathsetmacro{\xpos}{\xpos+1.5*\xstep}
    \node[rotate=90] at (\xpos+0.25*\xstep, \ypos) [black] {delay};
    \draw[thick] (\xpos, \ypos-0.50*\ystep) rectangle (\xpos+0.50*\xstep, \ypos+0.50*\ystep);
    \draw[black, thick, ->] (\xpos+0.5*\xstep, \ypos) -- (\xpos+1.0*\xstep, \ypos);
    % Comb
    \pgfmathsetmacro{\xpos}{\xpos+1.0*\xstep}
    \node at (\xpos+0.50*\xstep, \ypos) [black] {$H_\text{cav}$};
    \draw[thick] (\xpos, \ypos-0.50*\ystep) rectangle (\xpos+1.0*\xstep, \ypos+0.50*\ystep);
    \draw[black, thick, -] (\xpos+1.0*\xstep, \ypos) -- (\xpos+1.5*\xstep, \ypos);
    %
    % Open feedback
    \pgfmathsetmacro{\xpos}{\xpos+1.5*\xstep}
    \pgfmathsetmacro{\xposgbox}{\xpos}
    \draw[black, thick, -] (\xpos, \ypos) -- (\xpos+0.75*\xstep, \ypos) -- (\xpos+1.25*\xstep, \ypos+0.25*\ystep)  node[draw,circle,fill,scale=0.05*\size] {};
    \node at (\xpos+1.00*\xstep, \ypos+0.50*\ystep) [black,font=\scriptsize]  {open fb};
    \draw[black, thick, ->] (\xpos+1.25*\xstep, \ypos) node[draw,circle,fill,scale=0.05*\size] {} -- (\xpos+2.0*\xstep, \ypos);
    % Mod
    \pgfmathsetmacro{\xpos}{\xpos+2.0*\xstep}
    \node[rotate=90]  at (\xpos+0.25*\xstep, \ypos) [black] {mod};
    \draw[thick] (\xpos, \ypos-0.50*\ystep) rectangle (\xpos+0.50*\xstep, \ypos+0.50*\ystep);
    \node at (\xpos+1.0*\xstep, \ypos) [black,anchor=south] {$\Delta \vec{V}_\text{g}$};
    \node at (\xpos-0.50*\xstep, \ypos-0.25*\ystep) [black] {@$f_\text{r}$};
    \node at (\xpos+1.00*\xstep, \ypos-0.25*\ystep) [black] {@$f_\text{rf}$};
    \draw[black, thick, ->] (\xpos+0.5*\xstep, \ypos) -- (\xpos+2.0*\xstep, \ypos);
    % Sum fb+sp
    \pgfmathsetmacro{\xpos}{\xpos+2.0*\xstep}
    \pgfmathsetmacro{\xsumfbsp}{\xpos+0.5*\xstep}
    \node at (\xsumfbsp, \ypos) [black] {$\sum$};
    \draw[thick] (\xsumfbsp, \ypos) circle (0.5*\xstep);
    \node at (\xpos+0.0*\xstep, \ypos+0.0*\xstep) [black,anchor=south east] {$+$};
    \node at (\xpos+0.5*\xstep, \ypos+0.5*\ystep) [black,anchor=west] {$+$};
    \draw[black, thick, ->] (\xpos+1.0*\xstep, \ypos) -- (\xpos+1.5*\xstep, \ypos);
    % Gtx
    \pgfmathsetmacro{\xpos}{\xpos+1.5*\xstep}
    \node at (\xpos+0.375*\xstep, \ypos) [black] {$G_\text{tx}$};
    \draw[thick] (\xpos, \ypos-0.50*\ystep) -- (\xpos, \ypos+0.50*\ystep) -- (\xpos+\xstep,\ypos) -- (\xpos, \ypos-0.50*\ystep);
    \node at (\xpos+1.50*\xstep, \ypos+0.25*\ystep) [black] {$\vec{I}_\text{g}$};
    \draw[black, thick, ->] (\xpos+1.0*\xstep, \ypos) -- (\xpos+2.0*\xstep, \ypos);
    % Zgen
    \pgfmathsetmacro{\xpos}{\xpos+2.0*\xstep}
    \node at (\xpos+0.50*\xstep, \ypos) [black]  {$Z_\text{g}$};
    \draw[thick] (\xpos, \ypos-0.50*\ystep) rectangle (\xpos+1.0*\xstep, \ypos+0.50*\ystep);
    \node at (\xpos+2.0*\xstep, \ypos) [black,anchor=south] {$\vec{V}_\text{ind,g}$};
    \draw[black, thick, ->] (\xpos+1.0*\xstep, \ypos) -- (\xpos+2.0*\xstep, \ypos) -- (\xpos+2.0*\xstep, \ypos-0.5*\yrowspace+0.375*\ystep);
    %
    % Sum gen+beam
    \pgfmathsetmacro{\ypos}{\ypos-0.50*\yrowspace}
    \pgfmathsetmacro{\xpos}{\xpos+2.0*\xstep}
    \pgfmathsetmacro{\xsumgb}{\xpos}
    \node at (\xsumgb, \ypos) [black] {$\sum$};
    \draw[thick] (\xsumgb, \ypos) circle (0.5*\xstep);
    \node at (\xpos+0.0*\xstep, \ypos+0.5*\xstep) [black,anchor=south west] {$+$};
    \node at (\xpos+0.0*\xstep, \ypos-0.5*\ystep) [black,anchor=west] {$+$};
    \node at (\xpos+1.0*\xstep, \ypos) [black,anchor=south west] {$\vec{V}_\text{ind}$};
    \node at (\xpos+1.0*\xstep, \ypos) [black,anchor=north west] {@$f_\text{rf}$};
    \draw[black, thick, -] (\xpos+0.5*\xstep, \ypos) -- (\xpos+1.0*\xstep, \ypos) -- (\xpos+1.0*\xstep, \ypos-\yrowspace) -- (\xpos-15.75*\xstep, \ypos-\yrowspace);
    %
    % Zbeam
    \pgfmathsetmacro{\xpos}{\xpos-1.0*\xstep}
    \pgfmathsetmacro{\ypos}{\ypos-0.5*\yrowspace}
    \node at (\xpos-0.50*\xstep, \ypos) [black]  {$Z_\text{b}$};
    \draw[thick] (\xpos+0.0*\xstep, \ypos+0.50*\ystep) rectangle (\xpos-1.0*\xstep, \ypos-0.50*\ystep);
    \node at (\xpos+1.0*\xstep, \ypos) [black,anchor=north] {$\vec{V}_\text{ind,b}$};
    \draw[black, thick, ->] (\xpos+0.0*\xstep, \ypos) -- (\xpos+1.0*\xstep, \ypos) -- (\xpos+1.0*\xstep, \ypos+0.50*\yrowspace-0.375*\ystep);
    % LPF
    \pgfmathsetmacro{\xpos}{\xpos-2.0*\xstep}
    \node[rotate=90]  at (\xpos-0.25*\xstep, \ypos) [black] {LPF};
    \draw[thick] (\xpos, \ypos-0.50*\ystep) rectangle (\xpos-0.50*\xstep, \ypos+0.50*\ystep);
    \node at (\xpos+0.50*\xstep, \ypos+0.25*\ystep) [black]  {$\vec{I}_\text{b}$};
    \node at (\xpos+0.50*\xstep, \ypos-0.25*\ystep) [black]  {@$f_\text{rf}$};
    \draw[black, thick, ->] (\xpos+0.0*\xstep, \ypos) --  (\xpos+1.0*\xstep, \ypos);
    % deMod
    \pgfmathsetmacro{\xpos}{\xpos-1.0*\xstep}
    \node[rotate=90]  at (\xpos-0.25*\xstep, \ypos) [black] {demod};
    \draw[thick] (\xpos, \ypos-0.50*\ystep) rectangle (\xpos-0.50*\xstep, \ypos+0.50*\ystep);
    \node at (\xpos-1.00*\xstep, \ypos-0.25*\ystep) [black]  {DC};
    \draw[black, thick, ->] (\xpos+0.0*\xstep, \ypos) -- (\xpos+0.5*\xstep, \ypos);
    % dQ/dT
    \pgfmathsetmacro{\xpos}{\xpos-1.5*\xstep}
    \pgfmathsetmacro{\xposbbox}{\xpos-1.0*\xstep}
    \node at (\xpos-0.50*\xstep, \ypos) [black] {$\frac{\Delta Q}{\Delta t}$};
    \draw[thick] (\xpos, \ypos-0.50*\ystep) rectangle (\xpos-1.0*\xstep, \ypos+0.50*\ystep);
    \draw[black, thick, ->] (\xpos+0.0*\xstep, \ypos) -- (\xpos+1.0*\xstep, \ypos) ;
    %
    % Cav
    \pgfmathsetmacro{\xpos}{\xpos-7.00*\xstep}
    \node at (\xpos-1.00*\xstep, \ypos) [black] {cav};
    \draw[thick] (\xpos+0.00*\xstep, \ypos-0.26*\ystep) rectangle (\xpos-2.00*\xstep, \ypos+0.26*\ystep);
    \draw[thick,fill=white] (\xpos-1.50*\xstep, \ypos-0.23*\ystep) arc (225:315:1.4142*0.50*\xstep); % Bottom
    \draw[thick,fill=white] (\xpos-0.50*\xstep, \ypos+0.23*\ystep) arc ( 45:135:1.4142*0.50*\xstep); % Top
    \draw[thick,fill=white] (\xpos-0.25*\xstep, \ypos+0.00*\ystep) arc (. 0:180:1.4142*0.10*\xstep); % Antenna
    \node at (\xpos-0.25*\xstep, \ypos-0.5*\yrowspace) [black,anchor=south east] {$\vec{V}_\text{ind}$};
    \node at (\xpos-0.25*\xstep, \ypos-0.5*\yrowspace) [black,anchor=south west] {@$f_\text{rf}$};
    \draw[black, thick, -] (\xpos-0.25*\xstep, \ypos) -- (\xpos-0.25*\xstep, \ypos-0.5*\yrowspace) -- (\xposzero, \ypos-0.5*\yrowspace) -- (\xposzero, \yposzero);
    %
    % open drive
    \pgfmathsetmacro{\xpos}{0.5*\xsumantsp+0.5*\xsumfbsp}
    \pgfmathsetmacro{\ypos}{\yposzero+0.375*\yrowspace}
    \node at (\xpos-0.50*\xstep, \ypos+0.25*\ystep) [black,anchor=east] {$\vec{V}_\text{d}$};
    \node at (\xpos+0.50*\xstep, \ypos+0.25*\ystep) [black,anchor=west] {@$f_\text{fr}$};
    \draw[black, thick, <-] (\xsumantsp, \ypos-0.375*\yrowspace+0.375*\ystep) -- (\xsumantsp, \ypos) -- (\xpos-0.25*\xstep, \ypos) -- (\xpos+0.25*\xstep, \ypos+0.25*\ystep) node[draw,circle,fill,scale=0.05*\size] {} ;
    \node at (\xpos, \ypos-0.25*\ystep) [black,font=\scriptsize]  {open fb};
    \draw[black, thick, ->] (\xpos+0.25*\xstep, \ypos) node[draw,circle,fill,scale=0.05*\size] {} -- (\xsumfbsp, \ypos) -- (\xsumfbsp, \ypos-0.375*\yrowspace+0.375*\ystep); 
    %
    % llrf, gen, beam boxes
    \node at (\xposzero+0.50*\xstep, \yposzero+0.625*\yrowspace) [black,anchor=north west] {Feedback};
    \node at (\xposgbox+0.50*\xstep, \yposzero+0.625*\yrowspace) [black,anchor=north west] {Generator};
    \node at (\xposbbox+0.00*\xstep, \yposzero-1.375*\yrowspace) [black,anchor=south west] {Beam};
    \draw[dotted] (\xposzero+0.25*\xstep, \yposzero+0.675*\yrowspace) rectangle (\xposgbox,           \yposzero-1.0*\ystep); 
    \draw[dotted] (\xposgbox+0.25*\xstep, \yposzero+0.675*\yrowspace) rectangle (\xsumgb+0.75*\xstep, \yposzero-0.65*\ystep);
    \draw[dotted] (\xposbbox-0.25*\xstep, \yposzero-2.050*\ystep)     rectangle (\xsumgb+0.75*\xstep, \yposzero-1.425*\yrowspace);
  \end{tikzpicture}
    \caption{Schematic of the SPS one-turn delay feedback implementation in BLonD. The correction to the rf voltage along each turn is calculated by the cavity control from the difference of the cavity voltage (antenna), i.e.\ the sum of the beam- and generator-induced voltages, with the design voltage (set point).}
    \label{fig:SPS_cavitycontrol}
\end{figure*}
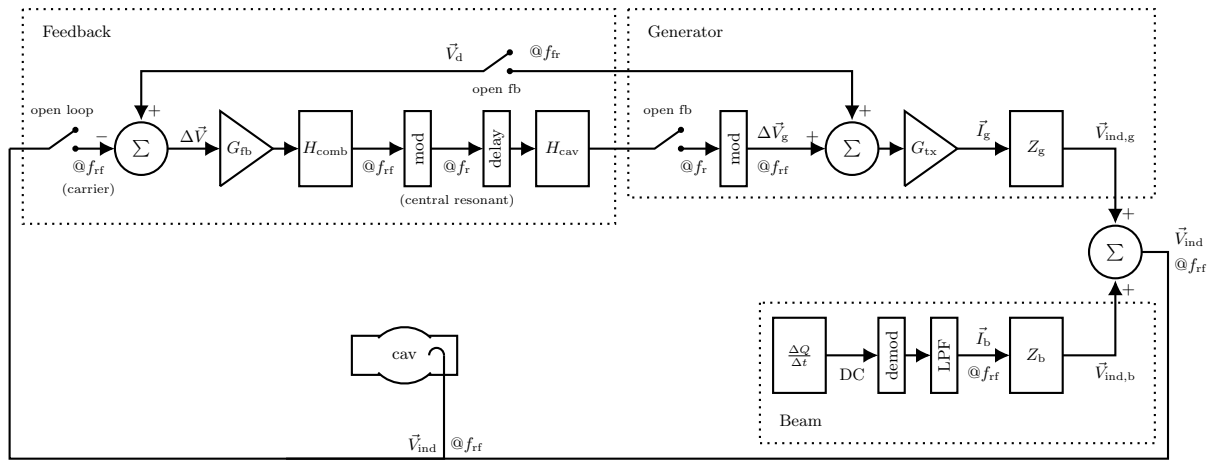

The accurate modeling of the feedback system of an accelerator is crucial for the realistic simulation of beam evolution, stability, and losses, as well as the assessment of the rf requirements of the machine. For example, the reduction of the bucket size resulting from the deviation of the design rf parameters, together with the modulation of the bunch-by-bunch positions resulting from the feedback loops used for beam-loading compensation can yield larger losses and reduce the machine performance. This becomes especially important for the bunch-to-bucket transfer between consecutive stages of an accelerator chain, between the PS-SPS and SPS-LHC~\cite{Schwarz:HB2018,Timko:HB2018-TUP1WA03,Medina:IPAC21lhclosses}. In the case of the High-Luminosity LHC (HL-LHC), the presence of large power transients at the head and tail of the high-intensity beam batches could result in a power demand beyond the capacity of the rf system; to understand their magnitude and dynamics, a realistic model of the full LHC rf cavity controller is needed~\cite{Medina:IPAC21cavcontrol}. The exact implementation of the cavity controller is machine-dependent, and models for the SPS (see \figref{fig:SPS_cavitycontrol}) and LHC controllers are presently available in BLonD.

Generally, a one-turn delay feedback system (OTFB) measures the cavity (antenna) voltage and applies the necessary correction in the following turn. In the SPS model, for instance, the antenna voltage $\vec{V}_{\mathrm{ind},(n)}$ is the sum of the beam- and generator-induced contributions ($\vec{V}_{\mathrm{ind,b},(n)}$, $\vec{V}_{\mathrm{ind,g},(n)}$, respectively)~\cite{Dome1976}:
\begin{equation} \label{eq:voltage-ind-ant}
    \vec{V}_{\mathrm{ind},(n)}(t) = \vec{V}_{\mathrm{ind,b},(n)}(t) + \vec{V}_{\mathrm{ind,g},(n)}(t),
\end{equation}
where $t$ is resolved typically on a bucket-by-bucket basis. The additional contribution from the reflected current is added, if applicable for the system. The vector notation refers to the use of complex voltage vectors. In its BLonD implementation, all these signals span one full turn are discretized at the rf frequency. As described in Sec.~\ref{sec:impedance}, the beam-induced voltage is the result of the cavity response to the rf beam current $\vec{I}_{\mathrm{b},(n)}(t)$ which is obtained from beam profile $\lambda_{(n)}(t)$ at $f_\mathrm{rf}$. Likewise, the generator component in \eqref{eq:voltage-ind-ant} is the result of the cavity response to the generator current $\vec{I}_{\mathrm{g},(n)}(t)$ according to the corresponding generator model (and its time evolution~\cite{Tuckmantel:1323893}). To compute the bunch-by-bunch correction to the rf voltage, the feedback system first computes the difference between the antenna voltage and the required design (or \emph{set point}) voltage $\vec{V}_{\mathrm{d},(n)}$,
\begin{equation}
    \Delta \vec{V}_{(n)}(t) = \vec{V}_{\mathrm{d},(n)} - \vec{V}_{\mathrm{ind},(n)}(t).
\end{equation}
This error signal is then processed (comb-filtered) 
to remove the beam loading effect by comparing it with the error in the \emph{previous} turn ($n-1$). 
The resulting signal constitutes an additional input to the generator drive, from which the corrected $\vec{I}_{\mathrm{g},(n)}(t)$ is obtained via the transmitter model. The corresponding generator-induced voltage $\vec{V}_{\mathrm{ind,g},(n)}$ will keep, in principle, the cavity voltage $\vec{V}_{\mathrm{ind},(n)}$ equal to the design voltage $\vec{V}_{\mathrm{d},(n)}$ on a bucket-by-bucket basis in the presence of beam loading. For rf power studies, the generator power is derived from these quantities.

For several cavities at the same harmonic in a given rf station, the total corrected voltage $\vec{V}_{\mathrm{corr},(n)}(t)$
%$\vec{V}_{\mathrm{sum},(n)}(t)$ 
is the sum of the cavity voltages $\vec{V}_{\mathrm{ind},(n)}(t)$ regulated by their corresponding feedback loop. The rf voltage seen by the beam in \eqref{eq:kick}, which is constant in amplitude and phase over a turn, is replaced for beam tracking by the following voltage that is modulated bucket by bucket:
\begin{equation} \label{eq:rfvoltage-feedback}
    V_{\mathrm{rf},(n)}(t)
    % Using Vcorr in terms of Vsum
    % = V_{\mathrm{d},(n)} V_{\mathrm{corr},(n)}(t) \sin\left(\omega_{\mathrm{rf},(n)} t + \varphi_{\mathrm{d},(n)} + \varphi_{\mathrm{corr},(n)}(t) \right),
    % Redefining Vcorr to simplify notation
    = V_{\mathrm{corr},(n)}(t) \sin\left(\omega_{\mathrm{rf},(n)} t + \varphi_{\mathrm{corr},(n)}(t) \right).
\end{equation}
%where $\vec{V}_{\mathrm{corr},(n)}$ is the total correction of the feedback systems relative to the design RF parameters,
% \begin{equation}
%     V_{\mathrm{corr},(n)}(t) = \frac{V_{\mathrm{sum},(n)}(t)}{V_{\mathrm{d},(n)}}
%     \qquad \text{and} \qquad
%     \varphi_{\mathrm{corr},(n)}(t) = \left( \frac{\pi}{2} - \varphi_{\mathrm{sum},(n)}(t) \right) -  \varphi_{\mathrm{d},(n)}.
% \end{equation}
For multi-harmonic rf systems, the contribution from additional rf harmonics should be added as described in \eqref{eq:total_voltage}.

The performance of the correction to the generator can be further increased by adding a feedforward loop on the beam branch, for example, as a FIR filter~\cite{Baudrenghien:2719232}. In the LHC and SPS models, the feedforward acts on the present-turn signals of the feedback output voltage and on the beam-induced voltage, respectively. Moreover, the feedback (including feedforward) loop can be part of or act together with other feedback loops such as analogue and digital feedback systems, as in the case of the LHC cavity controller, to provide additional corrections.

%%% DISTRIBUTIONS %%%
\section{\label{sec:distributions} Generation of particle distributions}

Once all the BLonD objects are initialized to treat the machine and impedance parameters, an initial particle distribution is needed to start the simulation. In this section, we describe several options that were included in the code to generate an initial particle distribution. The distribution can be generated either matched to the rf bucket or using an arbitrary density function. The density function in the longitudinal phase space is denoted $\distrib\left(\timecoord,\energycoord\right)$. The number of macro-particles to be generated is determined by the user, and each macro-particle corresponds to a fraction of the total beam current.

The criterion for a bunch of particles to be matched is that the density function $\distrib$ is a function of the hamiltonian $\hamiltonian$ (or the action $\action$). This implies that the particle density is uniform over an iso-hamoltonian curve, and therefore the particle distribution will remain stationary. Two routines were included to generate bunches matched to the rf bucket: the first one requires the density function $\distrib$ as a user input while the second one requires to input the line density $\lineden$.
\begin{figure*}[!t]
\centering
\subfloat[Using $\distrib$ as input. \label{fig:matching:distrib}]{
\begin{tikzpicture}[node distance=1.5cm, every node/.style={fill=white}, align=center]
%% Specification of nodes (position, etc.)
\node (machine) [input] {Machine and rf parameters};
\node (machinebis) [process, below of=machine] {$\potwellrf$ and $\eomfacdE$}; 
\node (hamiltonian) [process, below of=machinebis] {Hamiltonian $\hamiltonian$ \\or action $\action$}; 
\node (density) [input, below of=hamiltonian] {Density function \\$\distrib\left(\hamiltonian\text{ or }\action\right)$};
\node (lineden) [process, right of=density, xshift=3cm, yshift=-0.75cm] {Bunch line \\density $\lineden$};
\node (impedance) [input, above of=lineden] {Impedance $\impedance$};
\node (potdistort) [process, above of=impedance] {Potential well \\distortion $\potwellimp$};
\node (beam) [output, below of=density, yshift=-1cm] {Beam};
%% Specification of lines between nodes specified above
%% with aditional nodes for description 
\draw[->] (machine) -- (machinebis);
\draw[->] (machinebis) -- (hamiltonian);
\draw[->] (hamiltonian) -- (density);
\draw[->, dashed] (density) -- (lineden);
\draw[->, dashed] (lineden) -- (impedance);
\draw[->, dashed] (impedance) -- (potdistort);
\draw[->, dashed] (potdistort) -- (hamiltonian);
\draw[->] (density) --  node {Random seed} (beam);
\end{tikzpicture}
}%\\
\subfloat[Using $\lineden$ as input. \label{fig:matching:lineden}]{
\begin{tikzpicture}[node distance=1.5cm, every node/.style={fill=white}, align=center]
%% Specification of nodes (position, etc.)
\node (machine) [input] {Machine and rf parameters};
\node (machinebis) [process, below of=machine] {$\potwellrf$ and $\eomfacdE$}; 
\node (hamiltonian) [process, below of=machinebis] {Hamiltonian $\hamiltonian$}; 
\node (abel) [below of=hamiltonian] {Abel transform};
\node (lineden) [input, right of=density, xshift=3cm] {Bunch line \\density $\lineden$};
\node (impedance) [input, above of=lineden] {Impedance $\impedance$};
\node (potdistort) [process, above of=impedance] {Potential well \\distortion $\potwellimp$};
\node (density) [process, below of=abel] {Density function \\$\distrib\left(\hamiltonian\right)$};
\node (beam) [output, right of=density, xshift=3.5cm, minimum width=2cm] {Beam};
%% Specification of lines between nodes specified above
%% with aditional nodes for description 
\draw[->] (machine) -- (machinebis);
\draw[->] (machinebis) -- (hamiltonian);
%\draw[->] (hamiltonian) -- (density);
\draw[->, dashed] (hamiltonian) -- (lineden);
\draw[->, dashed] (lineden) -- (impedance);
\draw[->, dashed] (impedance) -- (potdistort);
\draw[->, dashed] (potdistort) -- (hamiltonian);
\draw      (lineden) -| (abel);
\draw      (hamiltonian.south) -| (abel);
\draw[->] (abel) -- (density);
\draw[->] (density) --  node {Random\\seed} (beam);
\end{tikzpicture}
}
\caption{Flowcharts of the routines used to generate a bunch of particles matched to the rf bucket in BLonD. The generation is an iterative process that minimizes the difference between the target emittance or bunch length asked by the user. The blue boxes represent the user input, the beige boxes are the parameters computed internally in the function, and the green boxes are the resulting beam objects used for tracking. The path in dashed arrows corresponds to the iterative loop for matching with collective effects. \label{fig:matching}}
\end{figure*}
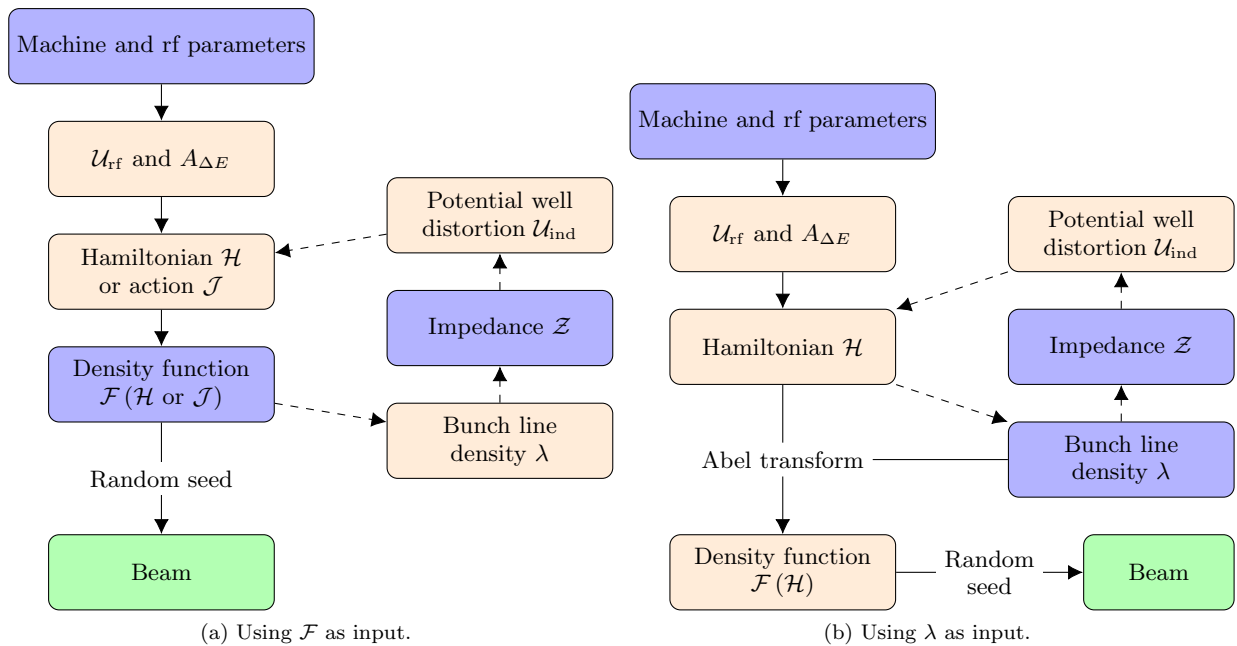

The routine using the density function $\distrib$ as an input is described in Fig.~\ref{fig:matching:distrib}. In this case, the hamiltonian is computed numerically on a grid in the $\left(\timecoord,\energycoord\right)$ phase space, using the input machine and rf parameters. The density function is then computed on the same grid, using the distribution function $\distrib$ chosen by the user from the ones given in the left column of Table~\ref{tab:matching}. If no impedance object was defined, a random particle distribution is directly sampled from the density function. If impedance sources are defined, the bunch is iteratively matched to take into account potential-well distortion. This method is particularly useful for studies scanning longitudinal emittance with a fixed density function. 

\begin{table}%[H] add [H] placement to break table across pages
 \caption{Density functions and their corresponding line density~\cite{Laclare:611596}. \label{tab:matching}}
 \begin{ruledtabular}
 \begin{tabular}{l|l|l}
    Name                & $\distrib\left(\hamiltonian\right)$                             & $\lineden\left(\timecoord\right)$                                                 \\ \hline
    Binomial~\footnote{$\mu=0$ waterbag, $\mu=1/2$ parabolic line density, $\mu=1$ parabolic in amplitude, $\distrib\left(\hamiltonian\geq\hamillen\right)=0$ and $\tau_{l}$ is the full bunch length}            & $\distrib_0\left(1-\frac{\hamiltonian}{\hamillen}\right)^{\mu}$ & $\lineden_0\left[1-4\left(\frac{\timecoord}{\bunchlen}\right)^2\right]^{\mu+1/2}$ \\
    Gaussian~\footnote{$\bunchlen=4\sigma$.}            & $\distrib_0\,e^{-2\frac{\hamiltonian}{\hamillen}}$              & $\lineden_0\,e^{-\left(\frac{4\timecoord}{\sqrt{2}\bunchlen}\right)^2}$
 \end{tabular}
 \end{ruledtabular}
 \end{table}

The second routine using the line density $\lineden$ as input is described in Fig.~\ref{fig:matching:lineden}. For this routine, the density in phase space is obtained using the inverse Abel transform~\cite{Krempl:Abel}. The inverse Abel transform was implemented numerically, and is applicable for cases where the potential well has only one minimum. Only a half of the bunch profile is required to compute the inverse Abel transform, the second half of the bunch profile results from the calculated density $\distrib$. This implies that in case of a non-symmetric potential well due to acceleration or collective effects, one half of the bunch profile is perfectly reproduced, and the measured and simulated profiles will only agree if the rf and impedance parameters are well known. In the presence of impedance sources, the matching consists of iteratively placing the bunch in the center of the distorted potential well till convergence, without changing the input line density. Once the density function $\distrib$ is obtained, the particle distribution is sampled randomly. This method is particularly useful for studies starting from bunch profiles identical to the ones obtained in measurements.

Both matching routines were extended for multi-bunch simulations. The methods implemented in BLonD to generate an initial particle distribution allowed to perform all kind of studies for all synchrotrons at CERN: starting with a matched distribution at any momentum during the acceleration ramp, with and without intensity effects, using the measured line density, etc. In addition, many routines to generate arbitrary density functions are implemented (e.g.\ bi-gaussian, coasting beam). Further options are continuously being implemented, to fulfill all needs encountered in the users' beam dynamics studies.

%%% OPTIMISATIONS %%%
\section{\label{sec:optimisation} Optimizations}

%\textcolor{red}{Kostis}
The demand for extensive and accurate longitudinal beam dynamics simulations is dictated by the ongoing upgrade projects, the studies for future machines as well as the operation of existing machines at CERN and other similar research facilities. A complete case study is typically composed of thousands of simulations, with the aim of identifying the set of parameters that optimally satisfies the target characteristics. Depending on the case, a single simulation can last up to several weeks. Therefore, optimizing the BLonD code in terms of run-time performance is crucial.

\begin{figure*}[t]
    \centering
    \includegraphics[width=0.7\textwidth]{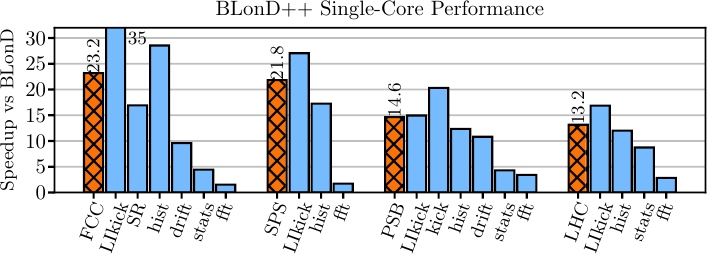}
    \caption{BLonD\texttt{++} speedup against the original Python-only BLonD implementation. All four testcases (FCC, SPS, PSB, LHC) were run using only one CPU core. The first bar of every group of bars shows the overall testcase speedup, while the remaining bars show the speedup of the most time-consuming code regions, which are `LIkick': linearly-interpolated energy kick, `SR': synchrotron radiation, `hist': line density calculation, `drift': drift calculation, `stats': statistics on beam distribution, `fft': Discrete Fast Fourier Transform.}
    \label{fig:opt-c-total-speedup}
\end{figure*}

\begin{figure*}[t]
    \centering
    \includegraphics[width=0.7\textwidth]{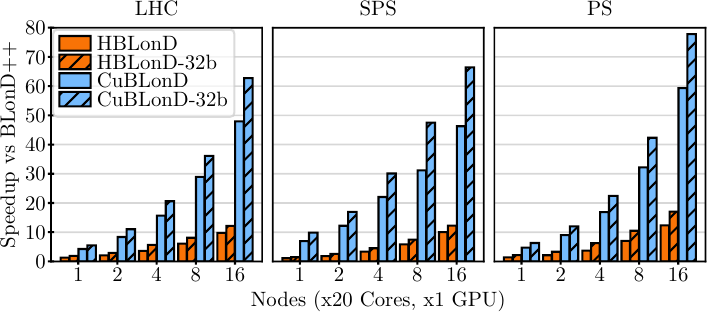}
    \caption{Multi-node performance and scalability of HBLonD and CuBLonD. From left to right, the bars in every group of bars correspond to the speedup compared to BLonD\texttt{++} of: a) HBLonD, b) HBLonD with reduced (32-bit) arithmetic precision, c) CuBLonD, and d) CuBLonD with reduced (32-bit) arithmetic precision.}
    \label{fig:opt-cublond}
\end{figure*}

BLonD was developed with performance in mind from the beginning. At first, the attention was focused on optimizing the serial or single-core performance of the code. For this purpose, the most time consuming code regions, the hotspots, were identified and translated to C\texttt{++}. C\texttt{++} is a lower-level, compiled language that generally allows for the development of performance critical software, and is supported by highly-efficient libraries such as the Boost library~\cite{schaling2011boost}, the Intel MKL library~\cite{intelmkl}, and the VDT library~\cite{Piparo_2014_vdt}. After analyzing and resolving a series of performance limitations in the hotspot regions, BLonD\texttt{++} was able to execute a previously day long simulation in 80~minutes~\cite{Iliakis:2018}. In Fig.~\ref{fig:opt-c-total-speedup}, we can see the single-core speedup of BLonD\texttt{++} compared to the original Python-only BLonD implementation, in four testcases that are well-representative of typical BLonD workloads. The first bar of every group shows the overall speedup while the remaining bars in each group show the speedup per accelerated code region. Since different testcases require a different configuration in terms of simulated macro-particles, number of beam slices, and simulation iterations, the per-testcase speedup ranges from 13.2$\times$, in the LHC testcase, to 23.2$\times$, in the FCC testcase. In addition, the most time-consuming code regions were parallelized using the OpenMP framework~\cite{dagum1998openmp}, enabling BLonD to scale vertically, i.e.\ within the boundaries of a single computing node. 

BLonD workloads comprise a scientifically and computationally challenging task. These workloads are typically inherently parallel and fit naturally in a distributed-memory environment. To anticipate the continuous need for simulations with finer resolution, more realistic modeling and longer simulation intervals, we developed a distributed version of BLonD, called HBLonD~\cite{kiliakis-scale-out}, that can efficiently combine multiple, remote  computing nodes to simultaneously calculate a BLonD simulation. The Message Passing Interface (MPI)~\cite{Snir:1998}, was used to allow the remote nodes to communicate. State-of-the-art high-performance computing techniques, like dynamic load-balancing, mixed-parallelism, and approximate computing, contributed to the impressive scalability demonstrated by HBLonD. Being able to efficiently combine up-to more than 600 cores across 32 computing nodes, HBLonD can offer greater than 10-fold speedups compared to BLonD\texttt{++}, thus reducing the simulation time further by one order of magnitude. 

% \begin{figure*}[t]
%     \centering
%     \includegraphics[width=0.7\textwidth]{optimizations/strong_scaling_cpu_gpu-lhc-sps-ps.pdf}
%     \caption{Multi-node performance and scalability of HBLonD and CuBLonD. From left to right, the bars in every group of bars correspond to the speedup compared to BLonD++ of: a) HBLonD, b) HBLonD with reduced (32-bit) arithmetic precision, c) CuBLonD, and d) CuBLonD with reduced (32-bit) arithmetic precision.}
%     \label{fig:opt-cublond}
% \end{figure*}

Graphics processing units (GPUs), originally designed to render images shown in display devices, e.g.\ a computer monitor, have lately become the dominant platform for accelerating general-purpose, data-parallel workloads. The most time-consuming tasks involved in a typical BLonD simulation, which are the particle tracking and the induced voltage calculation, are good candidates for GPU acceleration. Therefore, we implemented a GPU accelerated version of BLonD using the CUDA programming language~\cite{kirk2007nvidia}, called CuBLonD~\cite{iliakis2022}, which demonstrated an additional five-fold speedup compared to the CPU only version. In Fig.~\ref{fig:opt-cublond}, we performed a weak-scaling, multi-node performance evaluation of HBLonD and CuBLonD against BLonD\texttt{++}. In weak-scaling experiments, the amount of work per computing node is kept constant as the number of nodes increases. The horizontal axis of Fig.~\ref{fig:opt-cublond} shows the number of computing nodes, ranging from one to 16. Each computing node contains either 20 cores or one GPU platform. From left to right, the four bars in every group of bars correspond to the measured speedup against BLonD\texttt{++} of: a) HBLonD, b) HBLonD using reduced (32-bit instead of 64-bit) arithmetic precision, c) CuBLonD, and d) CuBLonD using reduced (32-bit instead of 64-bit) arithmetic precision. As we can see, HBLonD can effectively combine multiple computing nodes to run a single BLonD simulation, demonstrating speedups ranging from 10$\times$ up-to 17$\times$ compared to the previous, single-node implementation of BLonD. Furthermore, CuBLonD using GPUs to accelerate the calculation of the most time-consuming regions, achieves speedups that range from 46x up to 78x compared to BLonD\texttt{++}.

Presently CuBLonD is distributed together with the main BLonD code. To conclude, BLonD is an all-round optimized simulator, that adopts state-of-art high performance computing standards, and is capable of effectively utilizing the compute resources of multiple processors and GPU cards.

\section{Benchmarks}

Since its original release, BLonD has been used by a great number of scientists in a wide range of applications. The trust in the BLonD suite and its predictions has been established through in-depth testing and benchmarking. The conducted benchmarks, including comparisons with analytical calculations, measurements from experiments run in synchrotrons, or comparisons against other tracking codes~\cite{Timko:2016}, are all showing sharp agreement.

In addition, every care has been taken throughout the code development to ensure that BLonD results can be compared to measurement as accurately as possible. To this end, the input distributions can be idealized (e.g.\ Gaussian, waterbag, etc.) or taken directly from measurements. Also, the control loop features in the code are designed to reproduce the measured beam behaviour in the presence of these loops.

Below we show a few examples of code-to-code comparisons and benchmarks against theory and measurements.

\subsection{Code-to-code comparison with the MuSiC code}
Here we give an example of benchmark between the BLonD and MuSiC \cite{MiglioratiMultibunch} codes and we compare the obtained results with an analytical formula in narrow-band assumption~\cite{Quartullo:IPAC2017-THPVA022,DaniloThesis}. 

For a resonator impedance with a quality factor $Q\gg 1$ and relatively low resonant frequency $f_{\mathrm{r}}$, the wakefield can couple multiple bunches or even the same bunch on multiple turns. If the resonant frequency $f_{r}$ is close to an integer multiple $p$ of the revolution frequency, then Robinson instability can be observed \cite{chao1993physics}. Supposing a Gaussian line density with rms bunch length $\sigma_t$, the analytical expression for the growth rate is~\cite{chao1993physics}
\begin{align}
    \frac{1}{\tau_\mathrm{a}} = \frac{\eta e^2N_{p}}{2E_dT_{\mathrm{rev}}^2\omega_{s}}\sum_{m=\pm1} \Big( & m\,(p\omega_{\mathrm{rev}}+m\omega_{s}) \times \nonumber \\ 
    & \times \textrm{Re}Z(p\omega_{\mathrm{rev}}+m\omega_{s})\,G_m(x) \Big),
    \label{eq:growthrate}
\end{align}
where $G_m(x)=2 e^{-x^2}I_m(x^2)/x^2$ is the form factor with $x=(p f_{\mathrm{rev}}+m f_{s})\sigma_t$ and $I_m$ is the modified Bessel function of the first kind.

In the studied example, the wakefield of a single bunch couples the bunch to itself over thousands of turns. The integer multiple is $p=2$ and the resonator parameters are $f_r=2f_\mathrm{rev}+f_s$, $Q=~5000$ and $R_\mathrm{s}=40$~k$\Omega$. In addition, $N_\mathrm{p}=4\times 10^{12}$~ppb, $E_\mathrm{d}=13$~GeV, $\eta=0.0217$, $T_\mathrm{rev}=2.1$ \si{\micro}s, $f_\mathrm{s}=264.1$~Hz. The RF system has $h=7$, $f_\mathrm{rf}=3.3$~MHz and $\hat{V}_{rf}=165$~kV, while $R_{d}=100$~m. The instability growth time computed with \eqref{eq:growthrate} is $\tau_{a}= 59.3$~ms for $\sigma_t\leq 3.3$~ns and the results from MuSiC and BLonD should converge to $\tau_{a}$ for short bunches (no Landau damping). 

The initial bunch spectrum with $\sigma_t=3.3$ ns decays at 200~MHz whereas the resonant impedance is negligible above 1~MHz. It is then not easy to choose in BLonD the bin size $\Delta t$ in time-domain, or equivalently the maximum frequency $f_{\mathrm{max}}=1/(2\Delta t)$ in frequency domain. In addition, the frequency step $\Delta f=1/t_{\mathrm{max}}$ for Fourier transforms is another important parameter, since the wakefield decays slowly and it is not evident how many turns to take into account. The time domain approach for induced voltage calculation is used in BLonD since the narrow-band resonator would require a very small frequency step, making simulations computationally heavy. Indeed, the resonator impedance is not perfectly resolved even choosing $\Delta f=70$~Hz, which corresponds to $t_{\mathrm{max}}=7000$ $T_\mathrm{rev}$. The MuSiC approach avoids all these difficulties since it operates without slices and the number of macro-particles $N_{M}$ is the only parameter to be studied.

\begin{figure}[!htb]
    \centering
    \includegraphics*[width=0.48\textwidth]{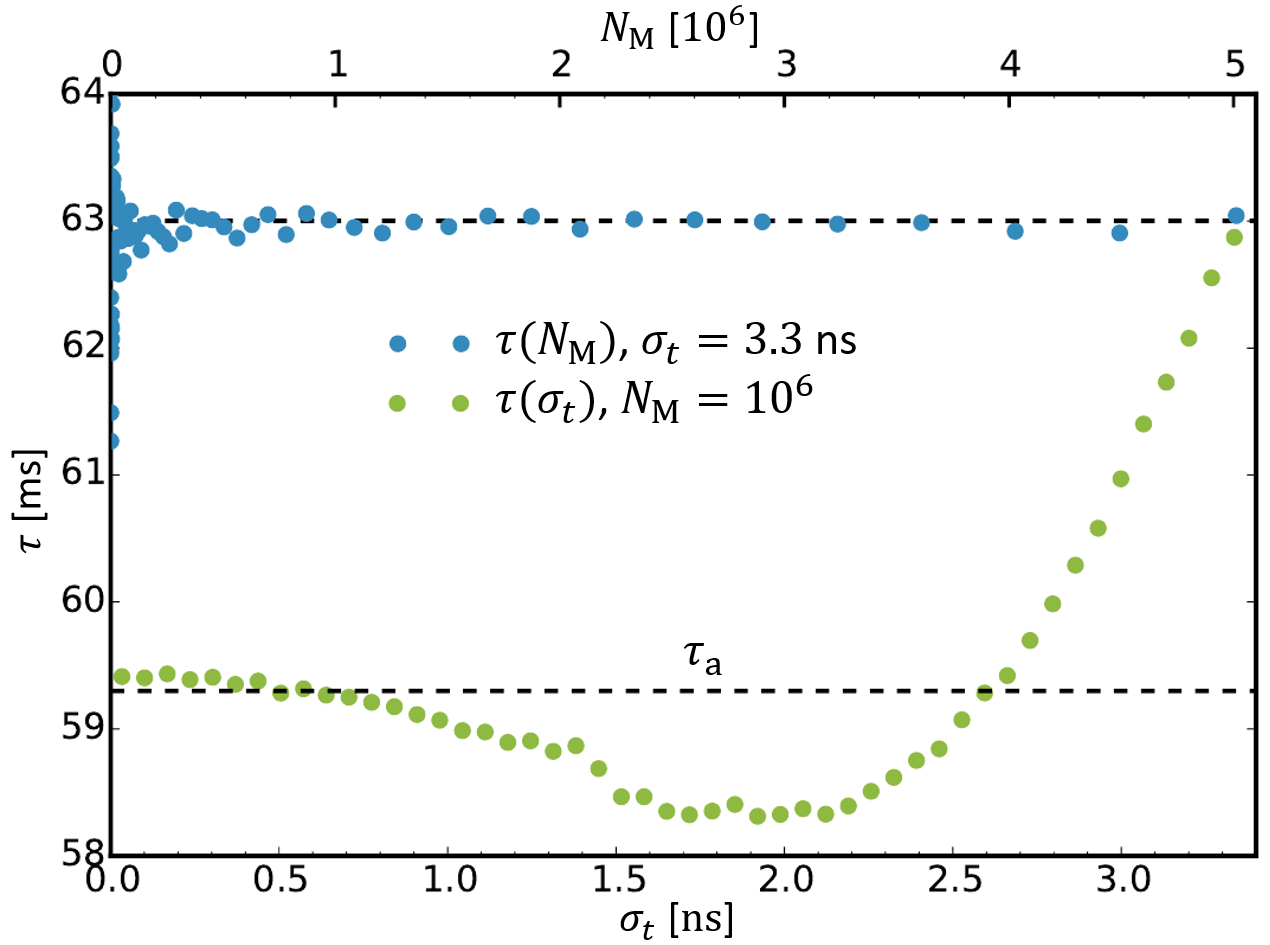}
    \caption{Instability growth time $\tau$ as a function of $N_{M}$ for $\sigma_t=3.3$~ns (blue) and as a function of $\sigma_t$ for $N_{M}=10^6$ (green) from MuSiC simulations. The dashed lines mark $\tau=\tau_{a}=59.3$~ms and $\tau=63$~ms.}
    \label{fig:growthrate}
\end{figure}

The instability growth time for $\sigma_t=3.3$ ns was examined in MuSiC as a function of $N_{M}$ (Fig.\ref{fig:growthrate}). When increasing $N_{M}$, convergence is observed (63.0~ms) but not to $\tau_a$, since the bunch is relatively long and Landau damping decreases the analytical growth rate. For lower bunch lengths, the growth time converges to $\tau_a$, proving the validity of the MuSiC algorithm. 

\begin{figure}[!htb]
    \centering
    \includegraphics*[width=0.48\textwidth]{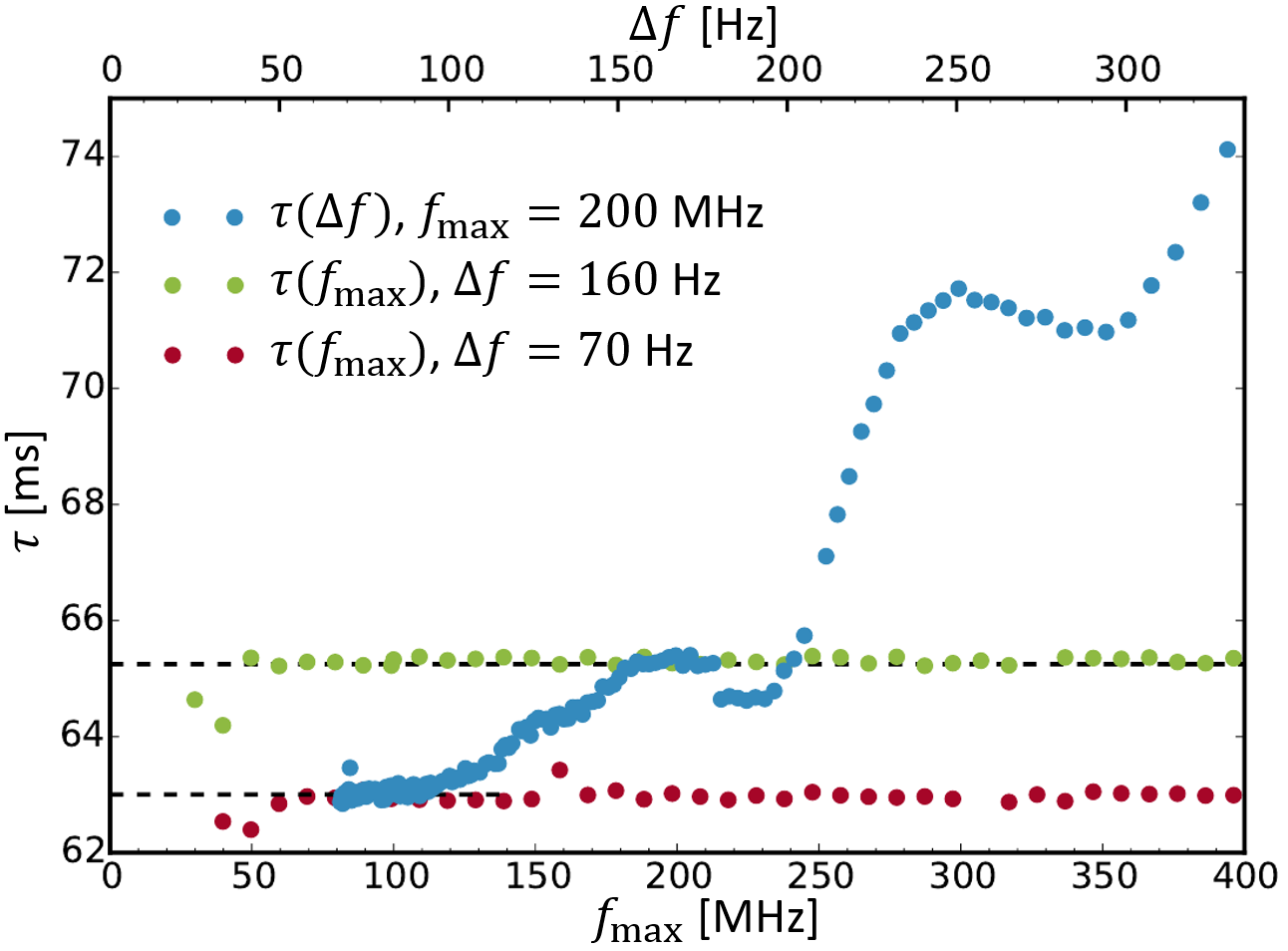}
    \caption{Instability growth time $\tau$ as a function of $\Delta f$ for $f_{\mathrm{max}}=200$~MHz (blue) and versus $f_{\mathrm{max}}$ for $\Delta f=160$~Hz (green) and $\Delta f=70$~Hz (red) from BLonD simulations ($\sigma_t=3.3$~ns, $N_{M}=10^6$). The dashed lines mark $\tau=63$~ms and $\tau=65.3$~ms. All the results shown are obtained through simulations in time-domain with $t_{\mathrm{max}}=1/\Delta f$ and $\Delta t=1/(2 f_{\mathrm{max}})$.}
    \label{fig:growthratedeltaf}
\end{figure}

Using $\sigma_t=3.3$~ns and $N_{M}=10^6$ in BLonD, the dependence of the growth time on $\Delta f$ and $f_{\mathrm{max}}$ was studied (Fig.\ref{fig:growthratedeltaf}). Choosing $f_{\mathrm{max}}=200$~MHz to properly cover the bunch spectrum, the growth time convergences to 63.0~ms for $\Delta f$ approaching 70~Hz, as expected from the MuSiC simulations. A scan of $f_{\mathrm{max}}$ for two given values of $\Delta f$ shows consistency of results, unless when $f_{\mathrm{max}}<50$~MHz, in which case the bunch spectrum is not properly covered and unreliable results are obtained. 

\subsection{Comparisons with the MELODY code}

\begin{figure}[tb!]
    \centering
    \includegraphics[width = 0.48\textwidth]{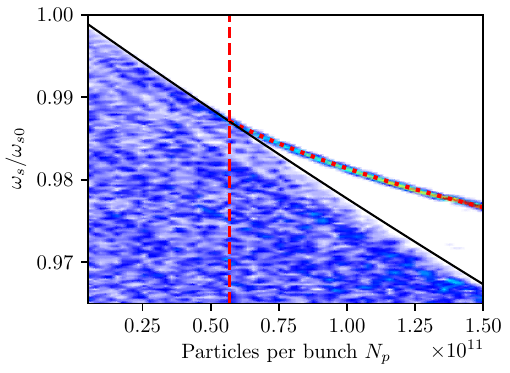}
    \includegraphics[width = 0.48\textwidth]{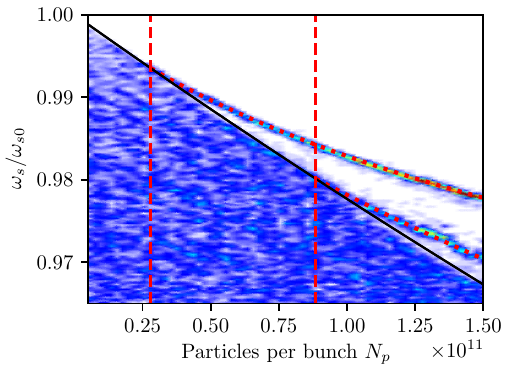}
    \caption{The real part of the normalized mode frequency found from macro-particle simulations using BLonD (blue color) and from MELODY (red dotted line) as a function of bunch intensity for a broad-band resonator impedance with Im$Z/n= 0.07 \;\Omega$, $Q=1$, $f_r = 4$~GHz (top) and $f_r = 8$~GHz (bottom). The maximum incoherent frequency obtained from MELODY is shown with black solid line. The dashed red lines indicate the LLD intensity thresholds. Plot from~\cite{Karpov2021}.}
    \label{fig:comp_modes_B_vs_M_bbr}
\end{figure}

Loss of Landau damping (LLD) in the longitudinal plane is an important intensity limitation in particle synchrotrons. New features were recently discovered in this field and summarized in Ref.~\cite{Karpov2021}. The new MELODY simulator (matrix equations for longitudinal beam dynamics)~\cite{MELODY} was written for numerical studies of LLD. For example, it was shown that the LLD threshold inversely proportional to the resonant frequency of the broad-band impedance model with quality factor of one. This is demonstrated in Fig.~\ref{fig:comp_modes_B_vs_M_bbr}, where we compare the emerged coherent modes that were obtained using the MELODY code and the BLonD code. The obtained results agree well.

\subsection{Benchmark with measurements}

% \begin{figure}
%     \centering
%     \subfloat[\label{fig:SPS_impedance_model}]
%     {\includegraphics[width=0.45\textwidth]{SPS_impedance_medium.png}}
%     \subfloat[\label{fig:meas_vs_BLonD_impedance}]
%     {\includegraphics[width=0.45\textwidth]{MD_287_projSpectra_forPaper.pdf}}
%     \caption{%\subref{fig:SPS_impedance_model} 
%     a) The SPS impedance model as of 2018~\cite{PhysRevAccelBeams.21.034401}, before an extensive impedance reduction campaign. 
%     b) Simulated and measured spectra of a debunching beam at SPS injection.}
% \end{figure}

\begin{figure*}
    \centering
    \includegraphics[height=0.3\textwidth]{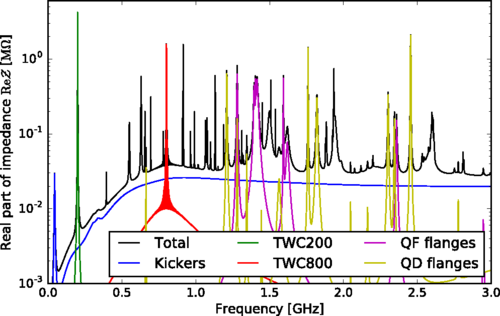}
    \includegraphics[height=0.3\textwidth]{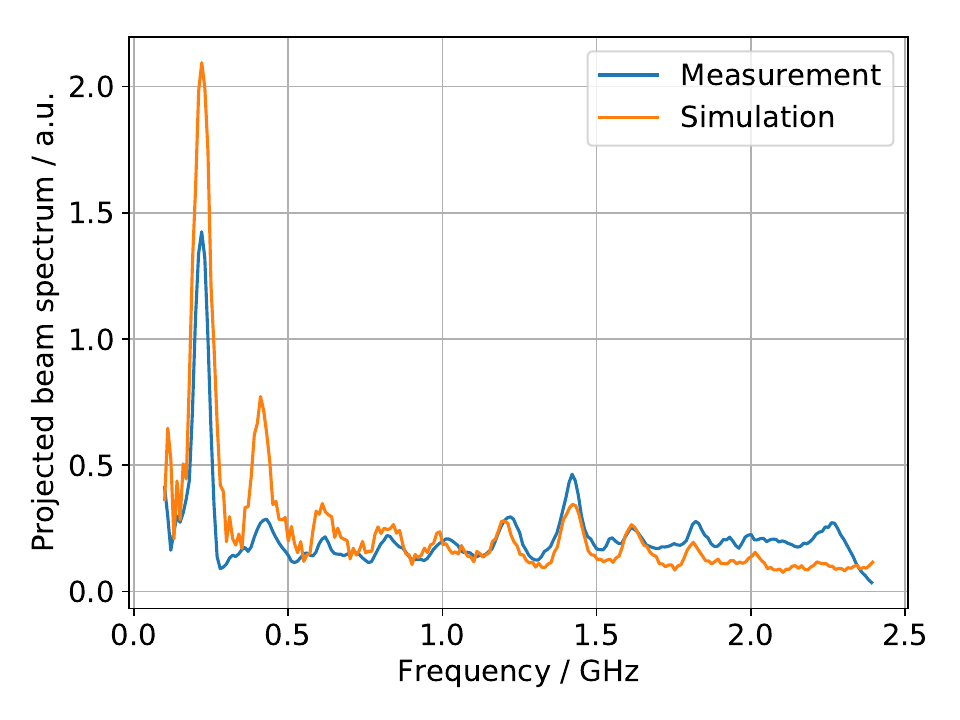}%\label{fig:meas_vs_BLonD_impedance}
    \caption{%\subref{fig:SPS_impedance_model} 
     \label{fig:SPS_impedance_model}
    The SPS impedance model (left) as of 2018~\cite{PhysRevAccelBeams.21.034401}, before an extensive impedance reduction campaign. 
    Simulated and measured spectra (right) of a debunching beam at SPS injection.}
\end{figure*}

As an application, we consider debunching measurements in the SPS~\cite{PhysRevAccelBeams.21.034401}, which are used to identify known and unknown sources of impedance in the machine. Long bunches are injected into the SPS with the rf system switched off, and the beam dynamics is entirely dictated by the machine impedance. Different impedance sources drive micro-wave instabilities, visible as a spatial modulation in the bunch profile.

The SPS impedance model is shown in \figref{fig:SPS_impedance_model} (left) and is dominated by the impedance of the travelling wave cavities (TWC) at 200~MHz and 800~MHz, their higher-order modes (HOMs), and vacuum flanges around 1.4~GHz. The SPS impedance sources range from broad-band, such as the kickers, to narrow-band, such as the 915~MHz HOM. In BLonD, the SPS impedance is modeled by over 200 resonators, several travelling wave cavities and impedance tables. 

The sum of the measured beam spectra during the first 500 revolutions is compared with the simulated spectrum in \figref{fig:SPS_impedance_model} (right). The initial phase-space distribution used in the simulation is based on measured profiles and tomographic reconstruction in the PS. Both spectra in \figref{fig:SPS_impedance_model} (right) have large amplitudes at 200~MHz and 1.4~GHz due to the significant impedances of the 200~MHz TWC and flanges, respectively.

\begin{figure*}[tb!]
    \centering
    \includegraphics[width = 0.48\textwidth]{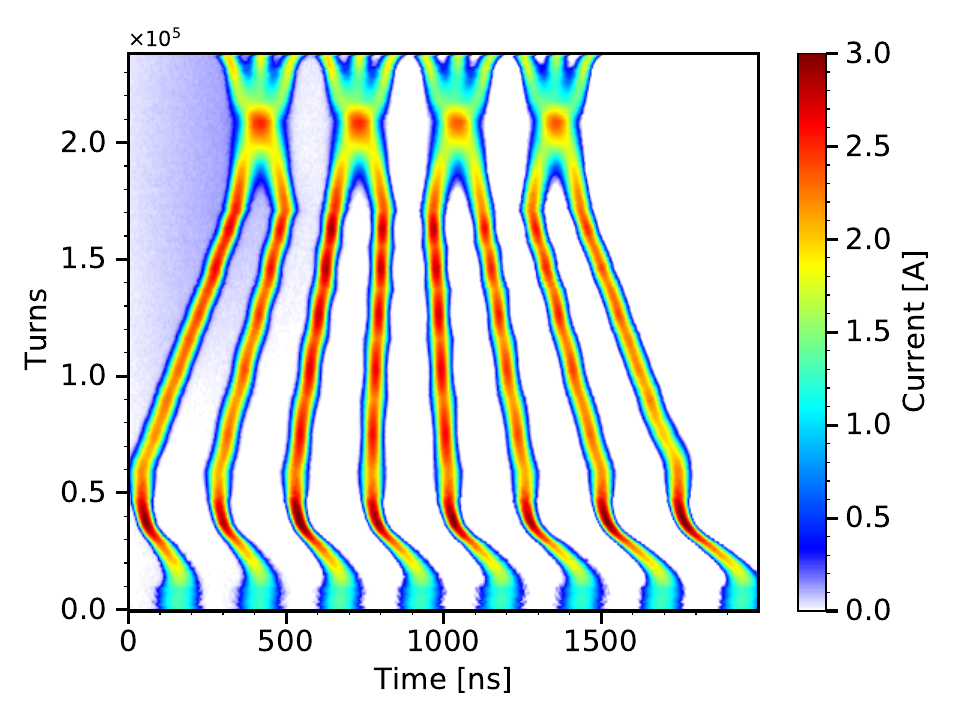}
    \includegraphics[width = 0.48\textwidth]{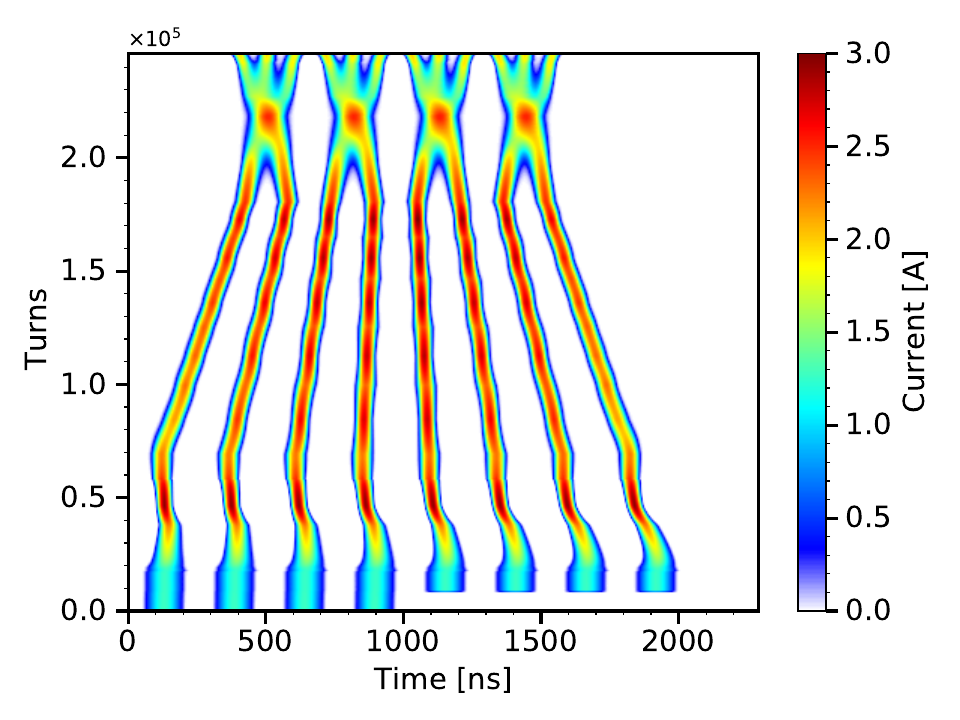}
    \caption{Comparison of the measured (left) and simulated (right) evolution of the bunch profiles during rf manipulations in the PS. The measured phase displacement of the beam between machine turns 10000 and 50000 is due to the signal delay with changing revolution frequency during acceleration and is not included in the simulation.}
    \label{fig:PS_manipulations}
\end{figure*}

Another example is the modeling of the rf manipulations in the PS, which are performed to get the nominal bunch spacing of 25~ns for LHC beams. These manipulations are made possible thanks to the large number of rf systems in the PS covering many rf harmonics. For the selected example shown in \figref{fig:PS_manipulations}, 8~bunches are injected from the PSB (two injections of 4~bunches) in $h=9$ and accelerated to an intermediate energy. Next, a batch compression is done to bring the beam to $h=14$ by passing through all intermediate harmonics. Bunches are then merged into $h=7$, before being finally being split in three to harmonic $h=21$. Note that a controlled longitudinal emittance blow-up is done right after the acceleration step with phase modulation of a high frequency rf system tuned at $h=436$. The nine harmonics can be programmed in BLonD to simulate the whole process. Beam loading effects were also included in the simulation as well as feedback systems. As seen on \figref{fig:PS_manipulations}, the simulation reproduces very well the measured beam evolution even for complex rf configurations.

%%% CONCLUSIONS %%%
\section{\label{sec:conclusions} Conclusions and Outlook}

The Beam Longitudinal Dynamics simulation suite BLonD has been presented. Its modular structure allows the user to build custom simulations modeling a range of physics phenomena, starting from basic cavity-beam interaction, over collective effects and a wide range of rf manipulations, to synchrotron radiation and feedback systems. The generation of matched particle distributions can be widely applied. Code optimizations on various levels and for a diverse range of hardware allow the user to make faster and more memory-efficient simulations.

In the future, it is planned to optimize the algorithm for the hamiltonian distribution functions that are used for the generation of particle distributions. The implementation of different algorithms for non-uniform binning and arbitrary impedance is on-going, too. Machine-specific global and local feedback models are constantly being added for various machines and are planned to be coupled.

BLonD is a constantly evolving and developing code base. Existing modules are updated and new modules are added to better tailor the studied phenomena or extend the current scope and functionality of the code. To ensure proper behavior and shielding against coding flaws, BLonD is supported by automatic unit-testing, integration testing and coverage analysis.

% \bibliography{bibliography}
%merlin.mbs apsrev4-1.bst 2010-07-25 4.21a (PWD, AO, DPC) hacked
%Control: key (0)
%Control: author (8) initials jnrlst
%Control: editor formatted (1) identically to author
%Control: production of article title (-1) disabled
%Control: page (0) single
%Control: year (1) truncated
%Control: production of eprint (0) enabled
%

\end{document}